\documentclass[12pt]{article}
\usepackage{amsmath,amssymb}
\newcommand{\be}{\begin{equation}}
\newcommand{\ee}{\end{equation}}
\newcommand{\bea}{\begin{eqnarray}}
\newcommand{\eea}{\end{eqnarray}}
\newcommand{\nn}{\nonumber \\}
\newcommand{\p}[1]{(\ref{#1})}
\newcommand{\lb}{\label}

\renewcommand{\u}{\underline}
\def\theequation{\arabic{section}.\arabic{equation}}
\begin{document}
\begin{titlepage}
\begin{flushright}
\end{flushright}
\vskip 0.6truecm

\begin{center}
{\Large\bf ${\cal N}{=}4$ Supersymmetric $d=1$ Sigma Models}
\vspace{0.2cm}

{\Large\bf on Group Manifolds} \vspace{1.5cm}

{\large\bf F. Delduc$\,{}^a$, E. Ivanov$\,{}^b$,}\\
\vspace{1cm}

{\it a)Univ Lyon, Ens de Lyon, Univ Claude Bernard, CNRS,\\ Laboratoire de Physique, F-69342 Lyon, France\\}
{\tt francois.delduc@ens-lyon.fr}
\vspace{0.3cm}

{\it b)Bogoliubov  Laboratory of Theoretical Physics, JINR,}\\
{\it 141980 Dubna, Moscow region, Russia} \\
{\tt eivanov@theor.jinr.ru}\\

\end{center}
\vspace{0.2cm}
\vskip 0.6truecm  \nopagebreak

\begin{abstract}
\noindent We construct manifestly ${\cal N}=4$ supersymmetric off-shell superfield actions for the HKT $d=1$ sigma
models on the group manifolds U(2) and SU(3), using the harmonic $d=1$ approach. The underlying $({\bf 4, 4, 0})$
and $({\bf 4, 4, 0})\oplus({\bf 4, 4, 0})$ multiplets are described, respectively, by one and two harmonic analytic superfields $q^+$
satisfying the appropriate nonlinear harmonic constraints. The invariant actions in both cases are bilinear in the
superfields. We present the corresponding superfield realizations of the U(2) and SU(3) isometries and show that in fact they are
enlarged to the products U(2)$\times$SU(2) and SU(3)$\times$U(2). We prove the corresponding invariances at both the superfield and component
levels and present the bosonic $d=1$ sigma model actions, as integral over $t$ in the U(2) case and over $t$ and ${\rm SU}(2)$ harmonics in the SU(3)
case. In the U(2) case we also give a detailed comparison with the general harmonic approach to HKT models and
establish a correspondence with a particular action of the off-shell nonlinear multiplet $({\bf 3, 4, 1})$.
A possible way of generalizing U(2) model to the matrix U($2n$) case is suggested.
\end{abstract}
\vspace{0.7cm}

\newpage

\end{titlepage}

\section{Introduction}
The ${\cal N}=4$ supersymmetric $d=1$ sigma models based on the multiplets with the off-shell content $({\bf 4, 4, 0})$
\footnote{In this notation \cite{PT}, the first two numerals stand for the number of physical bosonic and
fermionic fields and the third one for the number of auxiliary fields.} are known to lead to
the HKT (``Hyper-K\"ahler with torsion'') geometry in the bosonic target space and, as a particular case, the HK (``Hyper-K\"ahler'')
geometry \cite{hkt} - \cite{Hu}.  The intrinsic geometries of supersymmetric sigma models are displayed in the  clearest way in the appropriate
superfield formulations, with all supersymmetries being manifest and off-shell \cite{Zum} - \cite{HSS1}. For  HKT $d=1$ sigma models  with a target
space of real dimension $4n$ such a general formulation
was proposed in our paper \cite{DIN4}, where it was shown that they are described by $2n$ analytic harmonic superfields
$q^{+a}, a=1, \ldots 2n,$ (with the appropriate reality conditions)  subjected to the nonlinear harmonic constraint\footnote{Our notation is
the same as in \cite{DIN4}, \cite{IL}, \cite{gpr}. The reader can find there all the necessary information and further references related
to the $d=1$ harmonic superspace formalism.}
\bea
D^{++}q^{+ a} = {\cal L}^{+ 3a}(q^+, u^\pm)\,, \lb{L3M}
\eea
where ${\cal L}^{+ 3a}$ is an arbitrary analytic  function carrying the harmonic charge $+3$. The general
superfield action  is the following integral over the total ${\cal N}=4, d=1$ harmonic superspace,
\bea
S = -\frac18\,\int dt du d^4\theta  {\cal  L}(q^+, q^-, u^\pm)\,, \quad q^{- a} := D^{--}q^{+ a}\,. \lb{2ndprep}
\eea
The analytic function ${\cal L}^{+ 3a}(q^+, u)$ and the general function ${\cal L}(q^+, q^-, u^\pm)$ are two independent HKT ``prepotentials''
fully characterizing the given HKT geometry. The general expressions for the target metric and torsion were given and a few interesting
particular cases were discussed. It was shown that the general HK geometries arise for the simplest choice
\bea
{\cal L} = q^{+ a} q^{-b}\Omega_{ab}\,, \quad {\cal L}^{+3a}(q^+, u) = \frac{\partial {\cal L}^{+ 4}(q^+, u)}{\partial q^+_a}\,,  \lb{HK}
\eea
where $\Omega_{ab}=-\Omega_{ba}, \, \Omega_{ab}\Omega^{bc} = \delta^c_a\,,$ is a constant symplectic metric, while ${\cal L}^{+ 4}$ is the renowned
Hyper-K\"ahler potential \cite{hkHSS}, \cite{HSS}.  If  ${\cal L}^{+ 3a}(q^+, u)$ in \p{HK} is arbitrary, {\it i.e.}
is not subject to the second condition but ${\cal L}$ remains quadratic, the relevant geometries are strong HKT, that is, such that the corresponding
torsion is closed.  This class of HKT geometries is the same as in the $d=2$ ${\cal N}=(4,0)$ heterotic sigma models and the
analytic prepotential  ${\cal L}^{+ 3a}(q^+, u)$ coincides with the one introduced in \cite{dks}. The case of general
HKT potentials  ${\cal L}^{+ 3a}$ and ${\cal L}$ amounts to the ``weak'' HKT geometry,  in which there is no closedness condition
on the torsion two-form.  A more detailed geometric analysis of the two-potential formulation of the HKT geometry was undertaken
in a recent paper \cite{FIS}.

While for the HK manifolds a few concrete examples of ${\cal N}=4, d=1$ sigma models were presented (see, e.g., \cite{gpr}, \cite{bks} - \cite{ksch}),
not too many explicit examples of this kind are known for the HKT case. It still remains to find the precise form of the potentials for
the HKT manifolds known in the literature \footnote{It is comparatively easy \cite{FIS} to figure out these potentials  for one of the inhomogeneous strong HKT examples given in \cite{DV} (the
``Taub-NUT with torsion'' model).}. On the other hand, there
exists a wide class of homogeneous group-manifold strong HKT  metrics associated with those groups which admit a quaternionic
structure \cite{404, Joyce}. The full list of such groups  was given in \cite{404}, and it is tempting to explicitly construct the
relevant ${\cal N}=4$ HKT sigma models.  The simplest non-trivial series from this list are
\bea
({\rm a})\;\; {\rm SU}(2n)\times {\rm U}(1); \qquad ({\rm b})\;\; {\rm SU}(2n +1)\,, \lb{GMtwo}
\eea
with the dimensions $4n^2$ and $4n(n+1)$. Following reasoning of \cite{DIN4}, we can expect that the corresponding
superfield Lagrangians are always given by bilinears $\sim q^+ q^-$ and the whole non-triviality of one or another case
is encoded in the structure of the potential ${\cal L}^{+ 3 a}$.

In the present paper we construct the HKT sigma models for the simplest representatives of the series \p{GMtwo}, viz.
the group manifolds SU(2)$\times$U(1) and SU(3).  The first case is worked out in detail in sections 2 and 3. We start by giving the full
superfield formulation of the relevant sigma model and then consider the component bosonic limit of the superfield action.
The latter is the action of the $d=1$ sigma model with the target space $S^3\times S^1$, where
$S^3 \sim $ SU(2)$_L\times$SU(2)$_R/$SU(2)$_{diag}$ and SU(2)$_{diag}$ is none other than the standard harmonic SU(2) group.
The fact that the full symmetry of the model is the product  U(1)$\times $SU(2)$\times $SU(2) is non-trivial: this property  does not
immediately follow from the form  of the superfield action. We also gauge the U(1) symmetry of the model by
a non-propagating ${\cal N}=4$ gauge multiplet along the line of ref. \cite{gpr}. The outcome is a particular model
of the nonlinear $({\bf 3, 4, 1})$ multiplet with the $S^3$ sigma model in the bosonic sector. We present both superfield and
bosonic component actions for this case. In section 3  we perform the detailed comparison of the U(2) model with the general
harmonic approach to HKT models developed in \cite{DIN4} and find complete agreement. We explicitly construct the corresponding
closed torsion, as well as the triplet of quaternionic complex structures which are covariantly constant with respect to the relevant
Bismut connection.  The non-trivial symmetric Obata connection is also explicitly constructed, such that the complex structures are covariantly
constant with respect to it. The Riemann curvatures of both Bismut and Obata connections are shown to vanish.
In section 4 we proceed to the SU(3) case. It is much more complicated than the
U(2) case. This time, we deal with two independent $q^+$ superfields one of which is conveniently represented in the
$(N^{++}, \omega)$ basis, like in the previous case. We construct the corresponding SU(3) covariant superfield constraints
and show that the invariant superfield action is bilinear in the involved superfields, in full agreement with the general
assertion of ref. \cite{DIN4}. Further we pass to the bosonic limit and show that it reveals one more symmetry, U(2),
an essential part of which is the harmonic SU(2) as in the previous U(2) system. The extra U(2) is shown to commute with SU(3),
so the full symmetry of the model is SU(3)$\times$U(2). The relevant $d=1$ nonlinear sigma model is associated with he 8-dimensional coset  [SU(3)$\times$U(2)$]/$U(2)$_{diag}$.
We solve the bosonic sector of the initial superfield constraints,
find the bosonic action as an integral over $t$ and harmonics and explicitly do the harmonic integration in a part of this action.
In Appendix A, without performing the integration over harmonics, we independently demonstrate that the bosonic action possesses an    SU(3)$\times$U(2) invariance.
In the last section 5 we suggest a possible way of extending our considerations to generic $n$ in (\ref{GMtwo}a).
Some useful harmonic integrals are collected in Appendices B and C.

\setcounter{equation}{0}

\section{SU(2)$\times$U(1)  group manifold}
\subsection{Superfield constraints and action}
This case is described by two real harmonic analytic superfields $N^{++} = \widetilde{N^{++}}$ and $\omega = \widetilde{\omega}$ subjected to the harmonic
constraints\footnote{We use the same conventions as in \cite{DIN4}, \cite{gpr}. In particular, $\widetilde{\;}$ denotes the generalized conjugation \cite{HSS}, \cite{HSS1}
preserving the analytic
harmonic subspace  and  $D^{\pm\pm} = \partial^{\pm\pm} -2i \theta^\pm\bar\theta^\pm \partial_t + \theta^\pm \partial_{\theta^\mp} +  {\bar\theta}^\pm \partial_{{\bar\theta}^\mp}\,,
\widetilde{\theta^\pm} = {\bar\theta}^\pm\,, \quad \widetilde{{\bar\theta}^\pm} = -\theta^\pm\,$.}
\bea
&&D^{++}N^{++} + (N^{++})^2 = 0\,, \nn
&& D^{++}\omega - N^{++} = 0\,. \lb{Nomconstr}
\eea
This set of constraints is invariant under the following infinitesimal SU(2) transformations:
\bea
\delta N^{++} = \lambda^{++} - 2\lambda^{+-}N^{++} + \lambda^{--}(N^{++})^2\,, \quad \delta \omega = \lambda^{+-} - \lambda^{--}N^{++}\,, \lb{su2}
\eea
where
\be
\lambda^{\pm\pm} = \lambda^{(ik)}u^\pm_iu^\pm_k\,, \quad \lambda^{+-} = \lambda^{(ik)}u^+_iu^-_k\,,
\ee
and $\lambda^{(ik)}$ are constant symmetric parameters of SU(2). These harmonic projections satisfy the relations
\be
D^{++}\lambda^{++} = 0\,, \; D^{++}\lambda^{+-} = \lambda^{++}\,, \, D^{++}\lambda^{--} = 2 \lambda^{+-}\,.
\ee

The invariant sigma-model type action is written as an integral over the whole harmonic ${\cal N}=4$ superspace
\be
S_{su(2)} = -\frac{1}{4}\int dt du d^4\theta  \,N^{++}D^{--}\omega \lb{actNom}
\ee
(the particular normalization factor was chosen for further convenience). Using the fact that the integral of an analytic superfield over the full superspace is vanishing, as well
as the constraints \p{Nomconstr} together with the evident relation $D^{--}\lambda^{--} =0$,  it is easy to check
that the superfield action \p{actNom} is indeed invariant under the SU(2) transformations \p{su2} up to a total harmonic derivative
in the integrand. Though the action is bilinear in the involved superfields, in components it yields a nontrivial nonlinear SU(2)$\times$U(1) group
manifold sigma-model action. The extra U(1) acts as pure constant shifts of the superfield $\omega$
\be
\delta \omega = \lambda\,, \lb{u1}
\ee
and evidently leaves invariant both the constraints \p{Nomconstr} and the action \p{actNom}.

Combining the superfields $\omega$ and $N^{++}$ into yet another
analytic superfield $q^{+ a}$ as
\be q^{+ a} := u^{+ a}\omega - u^{-
a}N^{++}\,, \quad \omega = -(u^-\cdot q^+)\,, \; N^{++} = -(u^+\cdot
q^+)\,, \lb{Project}
\ee
the constraints \p{Nomconstr} and the action \p{actNom}
can be cast into the standard form \cite{DIN4}
\bea && D^{++} q^{+ a} = {\cal L}^{+ 3 a}\,, \quad {\cal L}^{+ 3 a} := u^{- a}(u^+\cdot q^+)^2\,, \lb{Constrqq} \\
&& S_{su2}
= -\frac{1}{8}\int dt du d^4\theta \, q^{+ a}D^{--} q^+_a\,.\lb{Actqq}
\eea
Based on the results of
\cite{DIN4}, this reformulation immediately implies that the bosonic
target space of the system under consideration is ``strong HKT'', that is a
HKT manifold with a closed torsion 3-form. At the component level,
this property is well known and is common to all group manifold HKT
models \cite{404}. In the present case we deal with the simplest
non-trivial example of such sigma model corresponding to the
four-dimensional target space SU(2)$\times$U(1). In what follows, it
will be more convenient to deal with the superfields $\omega$ and
$N^{++}\,$.

We also note that one more evident symmetry of the constraints \p{Nomconstr} and the actions \p{actNom},\p{Actqq} is the standard
harmonic SU(2)$_{H}$ realized on the harmonic variables as
\be
\delta_{H} u^\pm_i = \tau_{(i}^{\,\,k)}u^\pm_k\,. \lb{Hsu2}
\ee
It induces linear SU(2) rotations of all component fields with respect to their doublet indices. We will see later
that its presence ensures U(1)$\times$SU(2)$\times$SU(2) symmetry of the nonlinear sigma model appearing in the bosonic sector
of the action  \p{actNom}.

In fact, in the present case SU(2)$_{H}$ is a combination
of the standard automorphism group of ${\cal N}=4$ supersymmetry in the $q^{+a}$ representation and of the so called Pauli-G\"ursey SU(2)$_{PG}$, which commutes with supersymmetry
and  acts on the doublet index of the superfield $q^{+a}$. Such a non-uniqueness of the automorphism SU(2) group was mentioned in \cite{HSS1}.
It is with respect to such ``shifted'' SU(2)$_{H}$ that the superfield projections defined in \p{Project} are singlets. Though the action \p{Actqq}
(and \p{actNom}, up to a total harmonic derivative)
is formally invariant under SU(2)$_{PG}$ taken alone, the constraints \p{Constrqq} (or \p{Nomconstr}) are covariant only with respect
to the ``shifted'' SU(2)$_{H}$ .

\subsection{Bosonic component action and its symmetries}
In the limit where all fermionic fields are suppressed, the analytic superfields $N^{++}$ and $\omega$ have the following $\theta$
expansion
\be
N^{++} = n^{++} + \theta^+\bar\theta^+ \sigma\,, \quad \omega = \omega_0 + \theta^+\bar\theta^+\sigma^{-2}\,,\lb{expan}
\ee
where all component fields are defined on the coordinate set $(t, u^\pm_i)$. The harmonic superfield constraints \p{Nomconstr} imply
the following ones for the component fields
\bea
&& \partial^{++}n^{++} + (n^{++})^2 = 0\,, \nn
&& \partial^{++}\sigma - 2i\dot{n}{}^{++} + 2 n^{++}\sigma = 0\,, \nn
&& \partial^{++}\omega_0- n^{++} = 0\,, \nn
&& \partial^{++}\sigma^{-2} - 2i\dot{\omega}_0 - \sigma = 0\,. \lb{compconstr}
\eea
After rather simple manipulations which make use of the completeness condition for harmonics, $u^{+}_i u^-_k - u^{+}_k u^-_i = \varepsilon_{ik}$,
one solves all these equations, except for the last one, as
\bea
&& \omega_0 = b + \ln \left(1 + a^{+-}\right), \;\; n^{++} = \frac{a^{++}}{1 + a^{+-}}\,, \quad a^{+-} = a^{(ik)}u^+_i u^-_k\,, \;
a^{++} = a^{(ik)}u^+_{i} u^+_{k}\,, \nn
&& \sigma = \frac{1}{(1 + a^{+-})^2} \left\{c + 2i\left[\dot{a}^{+-} - a^{il}\dot{a}_l^{k}u^+_{(i}u^-_{k)}\right]\right\}, \lb{Sol}
\eea
where the fields $b, c$ and $a^{(ik)}$ depend only on $t$, but not on harmonics. After substituting these expressions into the
remaining equation for $\sigma^{-2}$ and taking the harmonic integral of both sides, we obtain a relation between the fields
$b(t)$ and $c(t)$
\be
c \int du\frac{1}{(1 + a^{+-})^2} + 2i\frac{d}{dt}\left[b + \int du \ln (1 + a^{+-}) - \int du \frac{1}{1 + a^{+-}}\right] = 0\,.\lb{BA}
\ee
Taking into account the last equation, we are left with a set of four $d=1$ fields $b(t), a^{(ik)}(t)$ parametrizing (as it will become clear soon)
the target SU(2)$\times$U(1) group manifold. The harmonic integrals entering \p{BA} can be explicitly carried out, giving us a simple expression for
the field $c(t)$ in terms of these basic fields. We dropped in \p{BA} the term
\be
-2i \int du\, \frac{(a^{il}\dot{a}^k_l)u^{+}_{(i}u^-_{k)}}{( 1 + a^{+-})^2}\,, \lb{Int1}
\ee
which is vanishing upon integration over harmonics.

The bosonic component action following from the superfield one \p{actNom} is
\be
S_{bos} = \frac{i}{2}\int dt du \left(n^{++}\dot{\sigma}^{-2} + \sigma \dot\omega_0 \right) :=  \int dt du {\cal L}(t,u) =
\int dt\,L(t) \,.\lb{compAct1}
\ee
After substituting the solutions \p{Sol} into ${\cal L}(t,u)$, integrating by parts with respect to $\partial_t$ and $\partial^{++}$,
as well as using the constraint for $\sigma^{-2}$, the Lagrangian ${\cal L}(t,u)$ can be presented in the following form
\bea
{\cal L}(t,u) &=& \dot{b}^2 + ic \frac{\dot{a}{}^{+-}}{( 1 + a^{+-})^3} - (\dot{a}^{+-})^2\frac{3 + a^{+-}}{( 1 + a^{+-})^3} \nn
&& +\, a^{il}\dot{a}_l^ku^+_{(i}u^-_{k)}\left[\frac{\dot b}{( 1 + a^{+-})^2} +\frac{2\dot{a}{}^{+-}}{( 1 + a^{+-})^3}\right]. \lb{bosAct2}
\eea

It remains to calculate the harmonic integrals in \p{BA} and \p{bosAct2}. Using the formulas \p{IntHarm1} - \p{IntHarm4}, after some algebra
\p{BA} can be reduced to
\be
c = -2i\,\dot{\tilde{b}} \left(1 + \frac{a^2}{2}\right), \quad \tilde{b} := b + \frac{1}{2}\ln \left(1 + \frac{a^2}{2}\right). \lb{AB1}
\ee
Next, as explained in Appendix A, the first term in the square bracket in \p{bosAct2} gives zero contribution to the harmonic integral
in \p{compAct1}. The harmonic integral of the second term can be represented as
$$
 \sim a^{il}\dot{a}_l^k \dot{a}^{mn} \int du\, \frac{u^+_{(m}u^-_{n)}\,u^+_{(i}u^-_{k)}}{(1 + a^{+-})^3}\,,
$$
and
$$
\int du\, \frac{u^+_{(m}u^-_{n)}\,u^+_{(i}u^-_{k)}}{(1 + a^{+-})^3} = a_{(mn)}\,a_{(ik)} f_1(a^2) + (\varepsilon_{mi}\varepsilon_{nk} +
\varepsilon_{ni}\varepsilon_{mk}) f_2(a^2)\,,
$$
which is the most general structure compatible with the SU(2) covariance. Substituting the second expression into the first one,
we immediately find that the latter is also vanishing. Thus only the first line in \p{bosAct2} contributes.
Using the formulas \p{J} - \p{J2}, it is easy to compute the remaining harmonic integrals and to present the final answer for
the Lagrangian ${L}(t)$ in \p{compAct1} as
\bea
L = (\dot{\tilde{b}})^2 + \frac{1}{2}\left[(\dot{a}\cdot \dot{a})\,\frac{1}{1 + \frac{a^2}{2}} - \frac{1}{2}(\dot{a}\cdot a)^2\,
\frac{1}{\left(1 + \frac{a^2}{2}\right)^2}\right].\lb{bosLagr}
\eea

Here, the field $\tilde{b}$ is invariant under the nonlinear realization of SU(2) acting on the second (sigma-model) piece of \p{bosLagr},
while $a_{(ik)}$ are just Goldstone fields supporting this nonlinear realization. Thus \p{bosLagr} describes a nonlinear $d=1$ sigma model
on the group manifold SU(2)$\times$U(1), parametrized by the $d=1$ fields $a_{(ik)}(t)$ and $\tilde{b}(t)\,$. The U(1) factor acts
as constant shifts of $\tilde{b}(t)\,$, while nonlinear SU(2) transformations of $a_{ik}$ can be found by considering the component-field
realization of the SU(2) symmetry originally defined on the superfields $\omega$ and $N^{++}\,$.

Starting from the transformation of $n^{++}$,
\be
\delta n^{++} = \lambda^{++} - 2\lambda^{+-} n^{++} + \lambda^{--}(n^{++})^2\,,
\ee
and using the first constraint in \p{compconstr}, one obtains an equation which determines $\delta a^{(ik)}$:
\be
\delta a^{++}(1 + a^{+-}) -\delta a^{+-} a^{++} = (1 + a^{+-})^2\lambda^{++} - 2\lambda^{+-} a^{++}(1+a^{+-}) + \lambda^{--}(a^{++})^2\,.
\ee
After some algebra, making use of the completeness relation for harmonics, one finds
\be
\delta a^{ik} = \lambda^{ik} + \frac{1}{2} (\lambda\cdot a) a^{(ik)} + \lambda^{(il}a_l^{k)}\,. \lb{Atransf}
\ee
It is easy to check that the Lie bracket of these transformations is again of the same form, i.e. we deal with a realization of SU(2)
on itself by left (or right) multiplications. The fields $a^{(ik)}(t)$ just provide a particular parametrization of the
SU(2) group element.

Analogously, starting from the SU(2) transformation of $\omega_0\,$,
\be
\delta \omega_0 = \lambda^{+-} - \lambda^{--}n^{++} = \lambda^{+-} -\lambda^{--}\,\frac{a^{++}}{1 + a^{+-}}\,,
\ee
and rewriting this variation in terms of the solution for $\omega_0$ in \p{Sol},
\be
\delta \omega_0 = \delta b + \frac{\delta a^{+-}}{1 + a^{+-}}\,,
\ee
one finds the SU(2) transformation law of the field $b(t)$:
\be
\delta b = -\frac{1}{2} (\lambda\cdot a)\,.
\ee
Now it is easy to show that the field $\tilde{b}(t)$ defined in \p{AB1} and entering the bosonic Lagrangian \p{bosLagr}
is indeed inert under SU(2)
\be
\delta \tilde{b} = 0\,.
\ee

We finally note that in fact the nonlinear sigma model  part of the Lagrangian \p{bosLagr} is invariant under two independent $SU(2)$ symmetries,
one being just \p{Atransf}, while another is the linearly realized SU(2) induced by the SU(2)$_H$ rotations of the harmonic variables \p{Hsu2}:
\be
\delta_H a^{ik} =  2 \tau^{(il}a_l^{\,k)}\,. \lb{Hsu2a}
\ee
This SU(2)$_H$ does not commute with \p{Atransf}. However, one can choose another basis for the generators of these two SU(2) groups,
such that the new SU(2) commutes with \p{Atransf}. This redefined SU(2) is given by the following transformations
\be
\delta a^{ik} = -\tau^{ik} - \frac{1}{2} (\tau\cdot a) a^{(ik)} + \tau^{(il}a_l^{k)}\,,  \lb{Atransf2}
\ee
where the signs chosen ensure that the relevant Lie bracket parameter is composed in the same
way as in the cases \p{Atransf} and \p{Hsu2a}. We denoted the transformation parameter in \p{Atransf2} by the same letter as in
\p{Hsu2a}, hoping that this will not give rise to confusion. The mutually commuting SU(2) transformations \p{Atransf} and
\p{Atransf2} in fact amount to the left and right shifts on the same SU(2) group manifold.

It is easy to see that the diagonal SU(2) subgroup in the product of these two commuting nonlinearly realized SU(2) is just SU(2)${}_H$ \p{Hsu2a},
so  the bosonic sigma-model in \p{bosLagr} describes a nonlinear realization of this product on the coset manifold
SU(2)$\times$ SU(2)$/$ SU(2)${}_H$, {\it i.e.}, the 3-sphere $S^3$.  The field $b$ transforms under the transformations \p{Atransf2} as
\be
\delta b = \frac{1}{2} (\tau\cdot a)\,,
\ee
while the redefined field $\tilde{b}$ is inert as in the case of $\lambda$-transformations. At the superfield level,
the $\tau$ transformations are generated by transformations of the same form as \p{su2}, modulo the common
sign minus before their r.h.s. and the appropriate addition of the linear harmonic-induced SU(2)$_H$
transformations of $N^{++}$ and $\omega$. Thus the full symmetry of the model under consideration, both on the superfield and the component
levels, is the direct product U(1)$\times$SU(2)$\times$SU(2)\,\footnote{Formally, the right SU(2) could be promoted to U(2), with U(1) being
realized by shifts undistinguishable from the left U(1) ones \p{u1}.}.

It is worth pointing out a crucial difference between the left and right SU(2) symmetries. The left one \p{su2} (as well as the constant shift \p{u1}) commutes with supersymmetry and
so defines the {\it triholomorphic} (or ``translational'' ) isometry of the model under consideration. It is realized as a subgroup of general
analytic reparametrizations of the superfields $q^{+ a}$. The right SU(2) involves the R-symmetry SU(2)$_H$ as its essential part and so {\it does not commute} with supersymmetry.
For what follows it is instructive to give the realization of the triholomorphic (left) SU(2) in terms of $q^{+ a}$:
\bea
&& \delta_\lambda q^{+ a} = u^{+ a}[\lambda^{+-} + \lambda^{--} (u^+\cdot q^+)] - u^{- a}[\lambda^{++} +2 \lambda^{+-} (u^+\cdot q^+) + \lambda^{--} (u^+\cdot q^+)^2], \lb{Tranq+} \\
&& \partial_{{+ b}}\,\delta_\lambda q^{+a} = - u^{+ a} u^+_b \lambda^{--} + 2 u^{-a}[ u^+_b \lambda^{+-} + u^+_b (u^+\cdot q^+)\lambda^{--}]\,, \lb{Tranq+2}
\eea
with $\partial_{{+ b}} := \frac{\partial}{\partial q^{+b}}\,.$

At this point, let us recall the existence of two distinguished connections on $S^3$ as a text-book example of a parallelizable manifold (see, e.g., \cite{BCZa}).
One of them is the standard Levi-Civita torsionless connection and it corresponds to treating $S^3$
as a symmetric Riemannian coset space SO(4)$/$SO(3). Another connection involves a closed torsion and has zero curvature, which corresponds to the identification of
$S^3$ with the non-symmetric SU(2) group manifold itself. This torsionful connection is what is called a ``Bismut connection'' (see \cite{FIS} and references therein) and it is the one
ensuring the SU(2)$\times$U(1) manifold to be HKT. Just with respect to the Bismut connection is the triplet of quaternionic complex structures covariantly constant.
In more detail, these geometric issues are discussed in section 3.

It is the appropriate place here to note that the 4-dimensional SO(4)$\times$ U(1)invariant  metric in \p{bosLagr} admits an equivalent representation as the conformally-flat
``Hopf manifold'' metric
\bea
\sim \frac{(\dot{x}_4)^2 + \dot{x}_m \dot{x}_m}{ (x_4 + c)^2 + x_m x_m}\,, \quad m = 1,2,3\,, \lb{Hopf}
\eea
where $c \neq 0$ is a constant (in fact it can be fixed as $c=1$ by a simple field redefinition). The SO(4) isometry acts as a rotation group of the four-dimensional Euclidean space
$\mathbb{R}^4 = (x_4 + c,  x_m)$, with the 3-dimensional rotations SO(3)$\subset$ SO(4) forming stability subgroup, while the U(1) (or R(1)) isometry as the
common constant rescaling of the $\mathbb{R}^4$ coordinates. The coset SO(4)$/$SO(3) transformations act inhomogeneously on the coordinates $x_m$, the same being true for
the transformation of $x_4$ under rescaling. In both cases the inhomogeneities are  proportional to $c$. It is rather straightforward to establish the equivalence relation between
the coordinate sets $(\tilde{b}(t), a^{(ik)}(t))$ and $(x_4(t), x_m(t))$. We prefer here to deal with the coordinates $(\tilde{b}(t), a^{(ik)}(t))$  because they manifest the product structure
$S^1\times S^3$ of the target space. The corresponding superfield description precisely matches the general case of the strong HKT ${\cal N}=4\,, \,d=1$
sigma models \cite{DIN4} and so can serve as a prototype for the analogous description of general group-manifold ${\cal N}=4\,, \,d=1$ models. It can be shown that the metric \p{Hopf}
naturally appears in an alternative description of the ${\cal N}=4$ U(2) model by a {\it linear} $q^{+a}$ multiplet, with vanishing ${\cal L}^{+3a}$ and more complicated ${\cal L}$. Such
a formulation and its relation to the one given here will be considered elsewhere (see Sect. 7 of \cite{FIS} for the relevant discussion).

\subsection{Reduction to nonlinear $({\bf 3, 4, 1})$ multiplet}
The last topic of the present section is the gauging of the U(1) symmetry \p{u1} along the line of ref. \cite{gpr}.
We promote the constant parameter $\lambda$ to an arbitrary analytic superfield parameter,
$\lambda \rightarrow \lambda(\zeta, u)$, $\omega^\prime = \omega + \lambda\,, N^{++}{}^\prime = N^{++}\,,$  and introduce two abelian harmonic connections
$V^{\pm\pm}$, $V^{\pm\pm}{}^\prime = V^{\pm\pm} - D^{\pm\pm}\lambda,$ such that $V^{++}$ is analytic,
$V^{++} = V^{++}(\zeta, u)$, while $V^{--}$ is related to $V^{++}$ by the covariant ``flatness'' condition,
\be
D^{++}V^{--} - D^{--}V^{++} = 0\,. \lb{flatn}
\ee
The constraints \p{Nomconstr} and the superfield action \p{actNom} are covariantized as
\bea
&&D^{++}N^{++} + (N^{++})^2 = 0\,, \nn
&& D^{++}\omega + V^{++} - N^{++} = 0\,,  \lb{Nomconstr1}
\eea
\be
S^{gauge}_{su(2)} = -\frac{1}{4}\int dt du d^4\theta  \,N^{++}(D^{--}\omega + V^{--}). \lb{actNom1}
\ee
The modified constraints and the action are, respectively, covariant and invariant under all rigid SU(2)
transformations discussed in this section. While checking this, an essential use of the flatness condition \p{flatn}
is needed.

Lets us show that the system so constructed describes the appropriate SU(2) invariant sigma model of the single
nonlinear multiplet $N^{++}$. To this end, we choose the ${\cal N}=4$ supersymmetric gauge
\be
\omega = 0\,.
\ee
In this gauge, the second constraint in \p{Nomconstr1} just amounts to identifying
\be
V^{++} = N^{++}\,,
\ee
while the action \p{actNom1} becomes
\be
S^{gauge}_{su(2)} = -\frac{1}{4}\int dt du d^4\theta  \,N^{++}V^{--}. \lb{actNom2}
\ee
Using \p{flatn}, $V^{--}$ can be expressed through $V^{++} = N^{++}$ by the well-know formula \cite{HSS1} involving
the harmonic distributions
\be
V^{--} = \int du_1 \frac{N^{++}(1)}{(u^+u^+_1)^2}\,,
\ee
where $V^{--}$ is taken at the harmonic ``point'' $u^\pm_i$, while $N^{++}$ in the r.h.s. is taken at the point $u^{\pm}_{1i}$.
Then the action \p{actNom2} can be written solely in terms of $N^{++}$ (in the central basis) as
\be
S^{gauge}_{su(2)} = -\frac{1}{4}\int dt d^4\theta  du_1 du_2\,N^{++}(1)\frac{1}{(u^+_1u^+_2)^2}N^{++}(2)\,. \lb{actNom2a}
\ee

The component bosonic  action  can be obtained in a simple form with the alternative choice of the Wess-Zumino gauge for $V^{\pm\pm}$
\be
V^{++} \rightarrow  2i\theta^+\bar\theta^+ A(t), \quad V^{--} \rightarrow  2i\theta^-\bar\theta^- A(t)\,, \quad  A^\prime = A + \dot\lambda(t)\,.
\lb{WZ}
\ee
In this gauge, the fourth of the component constraints \p{compconstr} is modified as
\be
\partial^{++}\sigma^{-2} - 2i\dot{\omega}_0 - \sigma + 2i A = 0\,, \lb{compconstr1}
\ee
while the bosonic action \p{compAct1} is modified as
\be
S_{bos} \rightarrow S_{bos}{}' = \frac{i}{2}\int dt du \left[n^{++}\dot{\sigma}^{-2} + \sigma (\dot\omega_0  - A) \right] :=
  \int dt du {\cal L}'(t,u) =
\int dt\,L'(t) \,.\lb{compAct2}
\ee
After some simple algebra, one finds that the only modification of the bosonic Lagrangian \p{bosLagr}
is the replacement
\be
\dot{\tilde{b}} \,\rightarrow \, \dot{\tilde{b}} - A\,.
\ee
Keeping in mind that under local U(1) transformation $\tilde{b}{\,}' = \tilde{b} + \lambda(t)$, one can always choose the gauge $\tilde{b}=0$,
after which the relevant bosonic Lagrangian will reduce to
\bea
L' = A^2 + \frac{1}{2}\left[(\dot{a}\cdot \dot{a})\,\frac{1}{1 + \frac{a^2}{2}} - \frac{1}{2}(\dot{a}\cdot a)^2\,
\frac{1}{\left(1 + \frac{a^2}{2}\right)^2}\right].\lb{bosLagr1}
\eea
The field $A$ proves to be just the auxiliary field of the nonlinear $({\bf 3, 4, 1})$ multiplet described
by the superfield $N^{++}$. It fully decouples in the bosonic action \p{bosLagr1} leaving us with the  $S^3$ non-linear
sigma model for 3 physical bosonic fields.

So, applying the general gauging procedure of ref. \cite{gpr} to the superfield Lagrangian \p{actNom} (or \p{Actqq}) of the U(2) HKT model, we recovered
a particular Lagrangian  of the nonlinear  $({\bf 3, 4, 1})$ multiplet, with the $S^3$ nonlinear sigma model
in the bosonic sector.\footnote{In the description through the standard (harmonic-independent) ${\cal N}=4, d=1$ superfields the nonlinear $({\bf 3, 4, 1})$
multiplet was studied in \cite{BeKr}. The case with the bosonic $S^3$ action \p{bosLagr1} was obtained as a particular case of the corresponding
general superfield action. Using a duality procedure at the component level, the bosonic SU(2)$\times$U(1) invariant action \p{bosLagr} was recovered through the dualization
of the auxiliary field $A$ into the physical scalar $b$.}

\setcounter{equation}{0}

\section{U(2) model as an example of HKT geometry in the harmonic approach}

Here we recover the metric and the torsion associated with the U(2) group manifold sigma  model directly from the general
geometric formalism of nonlinear $({\bf 4, 4, 0})$ multiplets pioneered in \cite{DIN4} and further elaborated on in \cite{FIS}.
Since we will be interested in the bosonic target geometry, we omit fermions and put $q^{+ a} = f^{+ a}$ altogether (in particular, in \p{Tranq+}, \p{Tranq+2}). The bosonic fields $\sigma$
and $\sigma^{-2}$ defined in \p{expan} are of use only while deriving the bosonic sigma model action from the superfield one. In the general
formalism of ref. \cite{DIN4} these additional quantities play no role, as they are eliminated there, from the very beginning,
by the harmonic superfield constraint.

The basic object of the geometric formalism is the ``bridge'' from the target $\lambda$ frame parametrized by the coordinates $f^{ + a}(t,u)$
and $f^{ - a}(t,u) := \partial^{--}f^{+ a}$ and harmonics $u^{\pm i}$ to the $\tau$ frame parametrized by the harmonic-independent coordinates, in our case
the $d=1$ fields $b(t)$ and $a^{ik}(t)$ defined in the previous subsection. The bridge is a $2\times 2$ complex matrix $M^{\underline b}_a$ subject to the
equation
\be
\partial^{++}M^{\underline b}_a + E^{+ 2 c}_a M_c^{\underline b} =0\,, \lb{Eq-bridge}
\ee
with
\be
 E^{+ 2 c}_a = \partial_{{+a}} {\cal L}^{+ 3 c} = -2 u^{-c}u^+_a (u^+\cdot f^+)\,. \lb{E+2}
\ee
The underlined doublet indices refer to the $\tau$ frame, and on them some harmonic-independent $\tau$ group acts. In the present case
the latter is realized as SU(2) rotations with the parameters $\lambda^a_b, \; \lambda^a_a =0$ (these parameters are the same as in \p{su2}, \p{Tranq+} and \p{Tranq+2}),
accompanied by some rescaling, see eqs. \p{IsomM}, \p{Tau}  below. On the non-underlined indices another realization of the same SU(2) is defined, as a particular isometry subgroup
of general harmonic analyticity-preserving target space reparametrizations.

Eq. \p{Eq-bridge} also implies that
\be
\partial^{--}M^{\underline b}_a + E^{- 2 c}_a M_c^{\underline b} =0\,, \lb{Eq-bridge1}
\ee
where the non-analytic connection $E^{- 2 c}_a$ is related to $E^{+ 2 c}_a$ by the ``harmonic flatness condition''
\be
\partial^{++}E^{- 2 c}_a = \partial^{--}E^{+ 2 c}_a + E^{+2 c}_d E^{-2 d}_a -  E^{+2 d}_a E^{-2 c}_d\,. \lb{E-2}
\ee
Fortunately, eq. \p{Eq-bridge} has the simple solution
\bea
M^{\underline b}_a = \delta_a^{\underline b}(1 + a^{+-}) - a^{++} u^{-{\underline b}}u^-_a\,, \quad (M^{-1})^c_{{\underline b}} = \frac{1}{1 + a^{+-}} \Big(\delta^c_{\underline b} +
\frac{a^{++}}{1 + a^{+-}} u^{-c} u^-_{\underline b}\Big). \lb{Bridge}
\eea
Now, the connection $E^{- 2 c}_a$ can be easily restored from \p{Eq-bridge1}:
\bea
E^{- 2 c}_a = -\Big[ \delta^c_b\, n^{--} + u^{- c}u^-_a \Big( n^{++}n^{--} - 2 \frac{a^{+-}}{ 1 + a^{+-}} \Big)\Big], \quad n^{\pm\pm} := (u^{\pm a} f^{\pm}_a) =
\frac{a^{\pm\pm}}{1 + a^{+-}}\,.
\eea
This expression can be checked to satisfy eq. \p{Eq-bridge1} \footnote{In fact, the same expression for $E^{- 2 c}_a$ can be directly derived by solving \p{Eq-bridge1}.}.
It is also straightforward to check that under the isometry \p{su2}, \p{Tranq+} and \p{Tranq+2}
the bridge $M^{\underline b}_a$ has the correct geometric transformation law (as anticipated above)
\be
\delta_\lambda \,M^{\underline b}_a = -\partial_{{+ a}}\,\delta_\lambda q^{+c}\,M^{\underline b}_c + \omega^{\underline b}_{\;\;{\underline c}}M^{\underline c}_a\,, \lb{IsomM}
\ee
where $\partial_{{+ a}}\,\delta_\lambda q^{+c}$ is given in \p{Tranq+2} and
\be
\omega^{\underline b}_{\;\;{\underline c}} = -\lambda^{({\underline b}}_{\;\;{\underline c})} + \frac12 \delta^{\underline b}_{\underline c}\,(\lambda\cdot a)\,,\lb{Tau}
\ee
are the  harmonic-independent parameters of the $\tau$ group the generical definition of which is given in \cite{DIN4}.

Having bridges at hand, we can calculate some ``semi-vierbeins'' which are building blocks of all geometric quantities in the approach of \cite{DIN4}. They transform the tangent
space objects to the world objects and, being specialized to the case under consideration,  are defined by
\footnote{We changed some  signs as compared to \cite{DIN4}.}
\bea
E^{+ \underline{a}}_{(ik)} = \partial_{(ik)}f^{+ a} M^{\underline{a}}_a = e^{\underline{i}\underline{a}}_{(ik)} u^+_{\underline i}\,, \quad E^{+ \underline{a}}_{4} =
\partial_{\tilde{b}}f^{+ a} M^{\underline{a}}_a = e^{\underline{i}\underline{a}}_{4} u^+_{\underline i}\,. \lb{DefViel}
\eea
Simple calculations yield
\bea
&& e^{\underline{i}\underline{a}}_{(ik)} = \frac12\,\big(\delta^{\underline i}_i \delta^{\underline a}_k + \delta^{\underline a}_i \delta^{\underline i}_k  \big)
- \frac{ a_{(ik)}}{2 + a^2}\,[ \varepsilon^{\underline{a}\underline{i}} + a^{(\underline{a}\underline{i})}]\,, \quad e^{\underline{i}\underline{a}}_{4} =
\varepsilon^{\underline{a}\underline{i}} + a^{(\underline{a}\underline{i})}\,, \nn
&& e_{\underline{i}\underline{a}}^{(ik)} = \frac12\,\big(\delta_{\underline i}^i \delta_{\underline a}^k + \delta_{\underline a}^i \delta_{\underline i}^k  \big)
- \frac12\, a^{(ik)}\,\varepsilon_{\underline{i}\underline{a}}\,, \quad e_{\underline{i}\underline{a}}^{4} = \frac{1}{2 + a^2}\, [ \varepsilon_{\underline{i}\underline{a}} +
 a_{(\underline{a}\underline{i})}]\,. \lb{Vielb}
\eea
The vierbein coefficients satisfy the standard orthogonality conditions. It can be checked that, under the triholomorphic SU(2) isometry, they transform
as covariant and contravariant tensors with respect to
non-underlined indices, with the infinitesimal parameters $\partial_{(ik)}\delta a^{(jl)}$, and as spinors with the parameters \p{Tau} with respect to the underlined doublet indices
 $\underline{a}, \underline{b}, \ldots$. On the indices $\underline{i}, \underline{k}, \ldots $ the standard automorphism SU(2) group acts.\\

Now we are prepared to compute the basic $\tau$ frame geometric objects of the U(2) model by specializing the general expressions of ref. \cite{DIN4}, \cite{FIS} to this case.
\vspace{0.3cm}

\noindent\underline{\it Metric}.  The target metric components are calculated by the general formula
\be
g_{ia \; jb} = e^{\underline{i}\underline{a}}_{ia }\,e^{\underline{j}\underline{b}}_{jb}\,\varepsilon_{\underline{i}\underline{j}} G_{[\underline{a}\, \underline{b}]}\,, \lb{Gen-g}
\ee
where
\be
G_{[\underline{a}\, \underline{b}]} = \int du {\cal F}_{[c\,d]}\,(M^{-1})^c_{\underline{a}} (M^{-1})^d_{\underline{b}}\,, \quad
{\cal F}_{[c\,d]}= \partial_{[+c}\partial_{-d]}\,{\cal L}\,. \lb{Gab}
\ee
In our case ${\cal L} = q^{+a} \varepsilon_{ab} q^{-b}$, so ${\cal F}_{[c\,d]} = \varepsilon_{cd}$. Using the harmonic integral formulas from Appendix B,
it is easy to find
\be
G_{[\underline{a}\, \underline{b}]}  = \varepsilon_{\underline{a}\, \underline{b}}\, \frac{1}{1 + \frac12 a^2}.
\ee
Substituting this into \p{Gen-g}, we find
\bea
&& g_{(ia) \; (jb)} = \frac{1}{2 + a^2}\Big[ \Big(\varepsilon_{ij}\varepsilon_{ab} + \varepsilon_{aj}\varepsilon_{ib}\Big) - \frac{1}{1 + \frac12 a^2}\; a_{(ia)} a_{(jb)}\Big], \nn
&& g_{4 \; (jb)} = 0\,, \quad g_{4 \; 4} = 2\,, \lb{Metr}
\eea
which precisely matches with the sigma-model Lagrangian \p{bosLagr}.

From now on, it will be convenient to pass to the 3-vector notation,
\be
a_{(ik)} = \frac{i}{\sqrt{2}}\,x_m\, (\sigma_m)_{ik}\,, \quad a_{(ik)}\,(\sigma_m)^{ik} =  -i\sqrt{2}\,x_m\,, \quad a^2 = x^2\,.\lb{TriplVect}
\ee
All world tensor indices $(ia)$ are replaced by the 3-vector ones by attaching the matrix factors $\frac{i}{\sqrt{2}} (\sigma_m)^{ia}\,$.
Then the metric takes the form \footnote{The new $d=1$ fields $x_m(t)$ should not be confused with those appearing in \p{Hopf}.}
\bea
g_{mn} = \frac{1}{1 + \frac12 x^2}\,\Big( \delta_{mn} - \frac12\frac{1}{1 + \frac12 x^2}\,x_m x_n \Big)\,, \quad
g^{mn} = \Big(1 + \frac12 x^2\Big)\,\Big(\delta_{mn} + \frac12\, x_m x_n\Big). \lb{xForm}
\eea

\vspace{0.3cm}

\noindent\underline{\it Torsion}. The torsion in the tangent space notation is expressed as
\be
C_{\u{i}\u{a} \,\u{k} \u{b}\, \u{l} \u{c}} =
\varepsilon_{\u{i}\u{l}}\,\nabla_{\u{k}\u{a}}\, G_{[\u{c} \u{b}]} +
\varepsilon_{\u{k}\u{l}}\,\nabla_{\u{i}\u{b}} \,G_{[\u{a} \u{c}]}
\ee
where
\be
\nabla_{\u{i}\u{a}}\, G_{[\u{c} \u{b}]}  = \int du \,(M^{-1})^{a}_{\u{a}}(M^{-1})^{c}_{\u{c}}(M^{-1})^{b}_{\u{b}}\,
\big(\partial_{- a}{\cal F}_{cb}\,u^-_{\u{i}} + {\cal D}_{+ a} {\cal F}_{cb}\,u^+_{\u{i}}\big) \lb{DefnablaG}
\ee
and
\bea
&& {\cal D}_{+ a} {\cal F}_{cb} = \nabla_{+ a}{\cal F}_{cb} + E^{-d}_{ac}{\cal F}_{db} + E^{-d}_{ab}{\cal F}_{cd}\,, \quad
 \nabla_{+ a} = \partial_{+a} + E^{-2 d}_a \partial_{-d}\,, \nn
&& {\cal D}^{++}  E^{-d}_{ac} =  E^{+d}_{ac}\,, \quad E^{+d}_{ab} =
\partial_{+a}\partial_{+b}\,{\cal L}^{+ 3 d}\,. \lb{DefE-daac}
\eea
In our case $G_{[\u{c} \u{b}]} := \varepsilon_{\u{c} \u{b}}G$ and
${\cal F}_{[c\,d]} = \varepsilon_{cd}$, so \p{DefnablaG} is simplified
to
\be
\nabla_{\u{i}\u{a}}\, G  = \int du
\,(M^{-1})^{a}_{\u{a}}(\det M)^{-1}E^-_a u^{+}_{\u{i}}\, \quad E^-_a := E^{-\,\,d}_{a d}\,.\lb{DefnablaGU2}
\ee
Next, $E^{+d}_{ab} = 2 u^{- d} u^+_a u^+_b$ and, by solving the harmonic equation in \p{DefE-daac}, we find
\be
E^{-d}_{ab} = - \frac43 \delta^d_{(a}u^-_{b)} + \varepsilon^{dc}u^-_{(c} u^+_{a} u^+_{b)}\,, \quad \Longrightarrow \quad E^-_a = -2u^-_a\,. \lb{E-a}
\ee
It is also easy to compute
\be
\det M = (1 + a^{+-})^2\,. \lb{detM}
\ee
Substituting \p{E-a} and \p{detM} in \p{DefnablaGU2}, we obtain
\bea
\nabla_{\u{i}\u{a}}\, G = -\int du \frac{1}{( 1 + a^{+-})^3} \big(\varepsilon_{\u{i}\u{a}} + 2 u^{+}_{(\u{i}}u^-_{\u{a})} \big). \nonumber
\eea
Calculating the harmonic integral with the help of eqs. \p{IntHarm5} and \p{J}  from Appendix B, we finally obtain
\bea
\nabla_{\u{i}\u{a}}\, G = -\frac{1}{(1 + \frac12 a^2)^2}\,\big[\varepsilon_{\u{i}\u{a}} + a_{(\u{i}\u{a})}\big] \lb{NablaFin}
\eea
and
\bea
C_{\u{i} \u{a} \,\u{k} \u{b}\, \u{l} \u{c}} = -\frac{1}{(1 + \frac12 a^2)^2}\,\big[\varepsilon_{\u{i}\u{l}}\varepsilon_{\u{c}\u{b}}(\varepsilon_{\u{k}\u{a}} + a_{\u{k}\u{a}}) +
\varepsilon_{\u{k}\u{l}}\varepsilon_{\u{a}\u{c}}(\varepsilon_{\u{i}\u{b}} + a_{\u{i}\u{b}}) \big]\,. \lb{TorTang}
\eea
It is easy to check the full antisymmetry of this expression with respect to permutation of the pairs of indices. Next we project this expression to the world-index representation
by contracting it with the ``semi-vierbeins'' \p{Vielb}. We find that all its components containing the index 4 vanish and only the projections on the 3-vector subspace are non-zero.
Substituting the triplet indices by the vector ones by the rule \p{TriplVect}, we finally obtain the only non-zero torsion component as
\be
C_{mns} = \sqrt{2}\frac{1}{(1 + \frac12 x^2)^2}\,  \varepsilon_{mns}\,. \lb{TorVect}
\ee
It is evidently closed as it should be for the group-manifold HKT models \cite{404}.
\vspace{0.3cm}

\noindent\underline{\it Bismut connection}. The Bismut connection on $4q$ dimensional HKT manifold is defined as
\be
\hat{\Gamma}^N_{MS} = \Gamma^N_{MS} + \frac12 g^{NP} C_{P MS}\,, \quad N, M,\ldots = 1,\dots\,, 4q\,, \lb{Bism}
\ee
where $\Gamma^N_{MS}$ is the standard symmetric Levi-Civita connection and $C_{PMS}$ is the torsion tensor. With respect to it, the triplet of the corresponding
quaternionic complex structures is covariantly constant. Since in our case (with $q=1$) $C_{4 mn} = 0$, the only non-vanishing components of the Bismut connection are
\be
\hat{\Gamma}^n_{ms} = \Gamma^n_{ms} + \frac1{\sqrt{2}}g^{np}\frac{1}{(1 + \frac12 x^2)^2}\,\varepsilon_{p m s}\,. \lb{BismU2}
\ee

The connection $\Gamma^n_{ms}$ for the metric defined in \p{xForm} is easily calculated to be
\be
\Gamma^n_{ms} = -\frac12 \frac{1}{1 + \frac12 x^2}\,\big( x_m\delta_{ns} + x_s\delta_{nm}\big)\,.\lb{LC}
\ee
The triplet of complex structures in the tangent space representation is given by
\be
(I_{A})^{\u{i}\u{a}}_{\u{j}\u{b}} = -i(\sigma_A)^{\u{i}}_{\u{j}} \delta^{\u{a}}_{\u{b}}\,,
\ee
where $\sigma_A\,, A=1,2,3\,,$ are Pauli matrices. Transforming it to the world indices by contracting with the vierbeins \p{Vielb} and  then passing to the indices $4, m$ through
the correspondence \p{TriplVect}, we explicitly find
\bea
(I_A)^4_4 &=& 0\,, \qquad (I_A)^4_m = \frac1{\sqrt{2}}\frac{1}{1 +\frac12 x^2}\Big( \delta_{mA} - \frac1{\sqrt{2}}\varepsilon_{Amp} x_p \Big), \nn
(I_A)^m_4 &=& -\sqrt{2}\Big(\delta_{Am} + \frac12\, x_A x_m - \frac1{\sqrt{2}}\, \varepsilon_{Amp}\, \,x_p\Big)\,, \lb{Compl1} \\
(I_A)^m_n &=& \varepsilon_{Amn} -\frac1{\sqrt{2}}\,\delta_{An}\,x_m + \frac1{\sqrt{2}} \,\frac{1}{1 +\frac12 x^2}\,x_n\Big(\delta_{Am}-\frac1{\sqrt{2}}\,\varepsilon_{Amp}\,x_p  +
\frac12\, x_A x_m \Big). \lb{Compl2}
\eea
Despite the somewhat involved form of these expressions, it can be checked that they form the algebra of imaginary quaternions and
possess the correct tensor transformation rules under the SU(2) isometry
$$
\delta_\lambda x_m = \lambda_m + \frac12 (\lambda\cdot x) x_m -\frac1{\sqrt{2}}\,\varepsilon_{mns}\,\lambda_n x_s\,.
$$
After some effort it can be also checked that they are covariantly constant with respect to $\hat{\Gamma}^m_{ns}$:
\bea
&& \partial_s (I_A)^4_m - \hat{\Gamma}_{sm}^p\,(I_A)^4_p = 0\,, \quad \partial_s (I_A)^m_4 + \hat{\Gamma}_{su}^m\,(I_A)^u_4 = 0\,, \nn
&& \partial_s (I_A)^m_n- \hat{\Gamma}_{sn}^p\,(I_A)^m_p
+ \hat{\Gamma}_{su}^m\, (I_A)^u_n = 0\,. \lb{CovConst}
\eea

One more important property of the Bismut connection $\hat{\Gamma}^m_{ns}$ (specific just for the group-manifold HKT models \cite{404}) is that its Riemann curvature vanishes
(the property pertinent to the parallelized manifolds \cite{BCZa}),
\bea
R^m_{\;\;nsp} (\hat{\Gamma}) = \partial_s\hat{\Gamma}^m_{pn} - \partial_p\hat{\Gamma}^m_{sn} + \hat{\Gamma}^m_{su}\hat{\Gamma}^u_{pn}- \hat{\Gamma}^m_{pu}\hat{\Gamma}^u_{sn} = 0\,. \lb{Curv1}
\eea
This property can also be directly checked using the explicit expression \p{BismU2} and \p{LC}.
\vspace{0.3cm}

\noindent\underline{\it Obata connection}. Besides the Bismut connection, one more important geometric object of HKT models is the Obata
connection \cite{Obata}. It is a symmetric connection with respect to which the triplet of complex structures  is still covariantly constant
(but not the metric, as distinct from the Levi-Civita connection). For HK manifolds it coincides with the Levi-Civita connection, but for HKT manifolds it does not.

In the harmonic approach to HKT geometry, Obata connection in the tangent space representation was defined in \cite{FIS}. It is the following deformation of the Bismut connection
\bea
\tilde{\Gamma}_{\u{i}\u{a} \,\u{k} \u{b}\, \u{l} \u{c}} = \Gamma_{\u{i}\u{a} \,\u{k} \u{b}\, \u{l} \u{c}}  + \frac12 \big(C_{\u{i}\u{a} \,\u{k} \u{b}\, \u{l} \u{c}} +
\Delta C_{\u{i}\u{a} \,\u{k} \u{b}\, \u{l} \u{c}}\big) = \hat{\Gamma}_{\u{i}\u{a} \,\u{k} \u{b}\, \u{l} \u{c}} +\frac12 \Delta C_{\u{i}\u{a} \,\u{k} \u{b}\, \u{l} \u{c}}\,, \lb{Obata}
\eea
where
\bea
\Delta C_{\u{i}\u{a} \,\u{k} \u{b}\, \u{l} \u{c}} = -2\varepsilon_{\u{i}\u{l}} \nabla_{\u{k}\u{c}}\,G_{[\u{a}\,\u{b}]}\,.
\eea
The full symmetry of $\tilde{\Gamma}_{\u{i}\u{a} \,\u{k} \u{b}\, \u{l} \u{c}}$ in the last two pairs of indices can be proved using the cyclic identity \cite{DIN4}
\bea
\nabla_{\u{k}\u{c}}\,G_{[\u{a}\,\u{b}]} + {\rm cycle}\,\big(\u{a}, \u{b}, \u{c} \big) =0\,. \nonumber
\eea

Now it is straightforward to compute the world-index form $\tilde{\Gamma}^M_{NS}, \; M, N, S = (4, m), (4, n), (4, s)\,,$ of the Obata connection for our model.
One should substitute $\nabla_{\u{k}\u{c}}\,G_{[\u{a}\,\u{b}]} \rightarrow \varepsilon_{\u{a}\u{b}}\nabla_{\u{k}\u{c}}\,G$, use eq. \p{NablaFin},  further contract \p{Obata}
with the proper vierbeins \p{Vielb} and finally pass to the 3-vector notation. The non-vanishing components of $\tilde{\Gamma}^M_{NS}$ prove to be
\bea
\tilde\Gamma^m_{ns} = \Gamma^m_{ns}\,, \quad \tilde\Gamma^m_{4n} = \tilde\Gamma^m_{n4} = \delta^m_n\,, \quad \tilde\Gamma^4_{ns} = -\frac12 g_{ns}\,,
\quad \tilde\Gamma^4_{44} =1\,, \lb{ObataComp}
\eea
where we took into account that $g^{44} = \frac12\,.$ Now it is a matter of somewhat tiring (though direct) computation to check the covariant constancy
of the complex structure $(I_A)^M_N$ defined in \p{Compl1}, \p{Compl2}:
\bea
\partial_p (I_A)^M_N - \tilde{\Gamma}_{p N}^S\,(I_A)^M_S + \tilde{\Gamma}_{p S}^M\, (I_A)^S_N = 0\,, \quad
\tilde{\Gamma}_{4 N}^S\,(I_A)^M_S - \tilde{\Gamma}_{4 S}^M (I_A)^S_N = 0\,.
\eea

It is also of interest to calculate the curvature of the Obata connection. Surprisingly, it turns out to vanish:
\bea
R^M_{\;\;NSP} (\tilde{\Gamma}) = 0\,.\lb{VanCurvOb}
\eea
According to ref. \cite{401a}, the vanishing of the curvature of the Obata connection for some HKT manifold signals that the three
corresponding complex structures are {\it simultaneously} integrable and so there exists a coordinate frame where all
these can be chosen constant. In the harmonic approach, this amounts to the conjecture \cite{FIS} that there exists a redefinition
of the original $q^{+}$ superfields such that they satisfy a {\it linear} harmonic constraint \footnote{${\cal N}=4$ HKT models
with linear harmonic constraints were addressed in \cite{IL} and, e.g., in \cite{FSmi}.}. This interesting issue will be studied
elsewhere (recall the related discussion in section 2.2 above).
\setcounter{equation}{0}

\section{SU(3) group manifold}

\subsection{Superfield consideration}

In this case, beside the superfields $N^{++}, \omega$ with four physical bosonic fields parametrizing SU(2)$\times$U(1),
we need one more analytic superfield in order to accommodate four extra bosonic fields which complete the SU(2)$\times$U(1) group manifold to
that of SU(3). The natural choice is the complex analytic superfield $(q^+, \bar{q}^+), \;\widetilde{(\bar{q}^+)} = - q^+\,$ where
``tilde'' stands for the generalized conjugation defined in \cite{HSS}, \cite{HSS1} and becoming the ordinary conjugation on the component fields.
Thus in the SU(3) case we operate with the following set
of analytic ${\cal N}=4, d=1$ superfields
\be
N^{++}\,, \;\;\omega\,, \;\; q^{+}\,, \;\; \bar{q}^+\,.
\ee
The extra fields appear as first terms in the harmonic expansion of $q^+, \bar{q}^+\,$, $q^+ = f^i u^{+}_i + \ldots\,, \;\;
\bar{q}^+ = \bar{f}^i u^{+}_i+ \ldots\,,$  $\bar{f}^i = \overline{(f_i)}\,$, i.e. they are doublets with respect to the automorphism SU(2).
The corresponding new constant group parameters are defined in a similar way
\be
\xi^\pm = \xi^iu^\pm_i\,, \quad \bar\xi^\pm = \bar\xi^iu^\pm_i\,, \quad \bar\xi^i = \overline{(\xi_i)}\,, \quad
\widetilde{(\bar{\xi}^\pm)} = -\xi^\pm\,.
\ee

Using the trial and error method, we have eventually found the following self-consistent set of the coset SU(3)/U(2) transformations for $q^+, \bar q^+, N^{++}$:
\bea
&&\delta q^+ = \xi^+ - \xi^-N^{++} + \alpha\, \xi^- q^+\bar{q}^+ + 2\alpha_0\, \bar\xi^-(q^+)^2\,, \nn
&&\delta \bar q^+ = \bar\xi^+ - \bar\xi^-N^{++} - \bar\alpha \bar\xi^- q^+\bar{q}^+ - 2\alpha_0 \xi^-(\bar q^+)^2\,, \nn
&& \delta N^{++} = \alpha\,(\xi^+ - \xi^-N^{++})\bar q^+ - \bar\alpha (\bar\xi^+ - \bar\xi^-N^{++}) q^+ -\alpha\bar\alpha
\left[\xi^-(\bar q^+)^2 q^+ + \bar\xi^-(q^+)^2 \bar q^+\right]. \lb{su3u2}
\eea
Here, $\alpha = \alpha_0 + i \alpha_1$ and $\alpha_1$ is, for the time being, an undetermined free real parameter. These transformations close on
the SU(2)$\times$U(1) ones:
\bea
(\delta_2\delta_1 - \delta_1\delta_2)q^+  := \delta_{br}q^+ &=&
 3i \alpha_0\,\phi_{br}\, q^+ - 2\alpha_0\left[(\lambda^{+-}_{br} - \lambda^{--}_{br}N^{++})q^+ + \alpha\,\lambda^{--}_{br} (q^+)^2\bar q^+ \right], \nn
\delta_{br}N^{++} &=& 2\alpha_0 \left[\lambda^{++}_{br} - 2 \lambda^{+-}_{br} N^{++}
+ \lambda^{--}_{br}(N^{++})^2 +
\alpha\bar\alpha\,\lambda^{--}_{br}(q^{+}\bar q^{+})^2\right],  \lb{u2}
\eea
where
\bea
\phi_{br} = -i[\xi_{(1)i}\bar{\xi}_{(2)}^i - (1 \leftrightarrow 2)],  \lambda^{\pm\pm}_{br} = \lambda^{ik}_{br}u^\pm_iu^\pm_k,\,
 \lambda^{+-}_{br} = \lambda^{ik}_{br}u^+_{(i}u^-_{k)},
\, \lambda^{(ik)}_{br} = [\xi_{(1)}^{(i}\bar\xi_{(2)}^{k)} - (1 \leftrightarrow 2)].
\eea
Note the necessary modification of the SU(2) transformation rule for $N^{++}$ as compared to the pure SU(2)$\times$U(1) case. Also note the
presence of a non-trivial U(1) phase  transformation in the closure on $q^+$ (the closure transformations of $\bar q^+$ can be obtained by
conjugation of $\delta_{br}q^{+}\,$). The non-zero real parameter $\alpha_0$ can be fixed at any value via a simultaneous rescaling of $q^+$ and of
the coset parameters $\xi^i\,$.

It remains to quote the corresponding transformation properties of the superfield $\omega$. Its coset SU(3)/U(2) transformation reads
\be
\delta \omega = \gamma\xi^-\bar q^+ - \bar\gamma \bar\xi^- q^+\,, \quad \gamma = \alpha_0 + i \gamma_1\,,\lb{omsu3}
\ee
where $\gamma_1$ is yet another undetermined real parameter. The closure of these transformations is the same as on the superfields $q^+, \bar q^+, N^{++}$,
with
\be
\delta_{br}\omega = -\gamma_1\phi_{br} +2\alpha_0\left[\lambda^{+-}_{br} - \lambda^{--}_{br}N^{++}
+ i (\alpha_1 - \gamma_1)\lambda^{--}_{br}(q^+\bar q^+) \right]. \lb{omu2}
\ee
The parameters $\alpha_1$ and $\gamma_1$ cannot be fixed from considering the closure of the above transformations: they form an $su(2)$ algebra
irrespective of the choice of these parameters. Surprisingly, they are fixed when constructing the invariant action. But before discussing this
issue, let us write down the relevant set of the harmonic constraints generalizing and extending \p{Nomconstr}. This set is as follows
\bea
&& D^{++}q^+ + N^{++}q^+ - \alpha (q^+)^2\bar q^+ = 0\,, \nn
&& D^{++}\bar q^+ + N^{++}\bar q^+ + \bar\alpha (\bar q^+)^2 q^+ = 0\,, \nn
&& D^{++}N^{++} + (N^{++})^2 + (\alpha\bar\alpha)\,(q^+\bar q^+)^2 = 0\,, \nn
&& D^{++}\omega - N^{++} + i(\alpha_1 - \gamma_1)\,q^+\bar q^+ = 0\,. \lb{su3constr}
\eea
One can check that the set \p{su3constr} is covariant under the SU(3) transformations \p{su3u2}, \p{omsu3} and \p{u2}, \p{omu2}.

It is interesting that the set of constraints \p{su3constr} for the choice $\alpha_1 = -\gamma_1$ can be cast in a simpler suggestive form.
Introducing
\be
\Phi^{++} := N^{++} - \alpha\,q^+\bar q^+\,, \quad \bar\Phi^{++} = N^{++} + \bar\alpha q^+\bar q^+\,,
\ee
we can rewrite \p{su3constr} as
\bea
&& (D^{++} + \Phi^{++})\,q^+ = 0\,, \quad D^{++}\Phi^{++} + (\Phi^{++})^2 = 0\,,\lb{hol1} \\
&& (D^{++}+ \bar\Phi^{++})\bar q^+ = 0\,, \quad D^{++}\bar\Phi^{++} + (\bar\Phi^{++})^2 = 0\,, \lb{anthol} \\
&& D^{++}\omega - \Phi^{++} -\bar\alpha\,q^+\bar q^+ = 0\,, \quad \mbox{or} \quad  D^{++}\omega - \bar\Phi^{++} +\alpha\,q^+\bar q^+ = 0\,. \lb{om2} 
\eea
The constraints \p{hol1} supplemented by the condition
\be
\Phi^{++} - \bar\Phi^{++} = -2\alpha_0\,q^+\bar q^+\,, \lb{phicond}
\ee
can be treated as the basic ones. The constraint \p{om2} serves just for expressing the superfield $\omega$ (up to the harmonic-independent part)
from the known $\Phi^{++}, q^+$ and the conjugated superfields. Note that the set $q^+, \Phi^{++}$ is closed under the SU(3)/U(2) transformations \p{su3u2}
\be
\delta \Phi^{++} = -2\alpha_0 \left(\bar \xi^+ q^+ - \bar\xi^-\Phi^{++}q^+\right), \quad \delta q^{+} = \xi^+ - \xi^- \Phi^{++}+ 2\alpha_0\bar{\xi}^-(q^+)^2
\lb{Phitr}
\ee
and hence under the U(2) ones as well:
\bea
\delta q^+ =  \alpha_0\,[{3i}\phi q^+ - 2\left(\lambda^{+-} - \lambda^{--}\Phi^{++}\right)q^+], \, \delta\Phi^{++}
=  2\alpha_0\,[\lambda^{++} - 2\lambda^{+-}\Phi^{++} + \lambda^{--}(\Phi^{++})^2].\lb{Phiu2}
\eea
Thus the set $(q^+, \Phi^{++}, u^\pm_i)$ can be interpreted as a kind of complex analytic subspace in the harmonic
extension of the target SU(3) group manifold (likewise, in the U(2) case the set $(N^{++}, u^\pm_i)$ can be
treated as some analytic subspace of the harmonic extension
of the U(2) group manifold).

The invariant action should be an extension of the action \p{actNom} and, following the general reasoning of ref. \cite{DIN4},
admit a formulation
in the full harmonic superspace as an expression  bilinear with respect to the superfields involved and linear
in the harmonic derivative $D^{--}\,$.
Combining \p{actNom} with $\bar q^+D^{--} q^+$, we find that requiring it to be invariant, up to a total harmonic derivative,
under the transformations
\p{su3u2} and \p{omsu3} (and, hence, under their closure), uniquely fixes the ratio of these two terms in such a way that
\be
S_{su(3)} = -\frac{1}{4}\int dt du d^4\theta  \left(N^{++}D^{--}\omega - 2\alpha_0\,\bar q^+ D^{--} q^+ \right),\lb{actsup2}
\ee
and, what is even more surprising, fixes the constants $\alpha_1$ and $\gamma_1$ in terms of the single
normalization constant $\alpha_0$ as
\be
\alpha = \alpha_0(1 \pm i\sqrt{3}), \quad \gamma = \alpha_0(1 \mp i\sqrt{3}), \lb{specalpha}
\ee
so that
\be
\alpha\bar\alpha = 4\alpha_0^2\,.\lb{specalpha1}
\ee
While checking the invariance, an essential use of the harmonic constraints \p{su3constr} was made.

It is instructive to give some technical details of the proof of the SU(3) invariance of the action \p{actsup2}. It will be enough to prove the invariance under the $\xi$
transformations, as the rest of SU(3) is contained in their closure. Moreover, it suffices to consider the holomorphic $\xi_i$ parts of these
transformations as the antiholomorphic ones $\sim \bar\xi^i$ are obtained through the tilde-conjugation. So we start from the transformations
\bea
&&\delta N^{++} = \alpha\big[\xi^+ - \xi^- \big(N^{++} + \bar\alpha\, q^+\bar q^+\big)\big] \bar q^+\,, \quad \delta \omega = \gamma\, \xi^- \bar q^+\,, \nn
&&\delta q^+ = \xi^+ - \xi^- \big(N^{++} - \alpha q^+\bar q^+\big)\,, \quad \delta \bar q^+  = -2\alpha_0\, \xi^- (\bar q^+)^2\,, \lb{Holomxi}
\eea
and the following ansatz for the superfield Lagrangian
\be
L = L_1 + \kappa L_2\,, \qquad L_1 = N^{++} D^{--}\omega\,, \quad L_2 = \bar q^+ D^{--} q^+\,,
\ee
where $\kappa$ is some real parameter. The variations of $L_1$ and $L_2$, up to total time and harmonic derivatives and upon using the harmonic constraints
\p{su3constr} along with
the property that the full superspace integral of an analytic expression vanishes,  are reduced to
\bea
\delta L_1 \; &\Rightarrow& \; -(\alpha + \gamma) \xi^- \bar q^+ D^{--} N^{++} + i\frac{\alpha}{2}(\alpha_1 - \gamma_1) \xi^- ({\bar q}^+)^2 D^{--}q^+\,, \nn
\delta L_2 \; &\Rightarrow& \; -\xi^- \bar q^+ D^{--} N^{++} + \big(\frac{\alpha}{2} - 2\alpha_0 \big)\xi^- ({\bar q}^+)^2 D^{--}q^+\,.
\eea
The superfield structures in these variations are independent, so the conditions for the vanishing of  $\delta L$  are
\bea
\kappa + \alpha + \gamma = 0\,, \quad i\frac{\alpha}{2}\,(\alpha_1 - \gamma_1) + \kappa\,\big(\frac{\alpha}{2} - 2\alpha_0 \big) =0\,.
\eea
The separate vanishing of the real and imaginary parts of these equations uniquely yields
\be
\alpha_1 = -\gamma_1\,, \quad \kappa = -2\alpha_0\,, \quad \alpha_1^2 = 3\alpha_0^2\,,
\ee
that precisely matches with \p{actsup2} - \p{specalpha1}.

It is also easy to directly prove the invariance of \p{actsup2} under the U(2)$\,\subset\,$SU(3) transformations and
under the transformations of the second U(2) commuting with SU(3) (the second SU(2) transformations are a fixed combination of the first SU(2) ones and
of those of the harmonic SU(2), while the second U(1) factor acts just as a constant shift of $\omega$, see eqs. \p{Au22} - \p{tau0} below).

\subsection{Bosonic limit and its symmetries}
Here we solve the purely bosonic part of the constraints \p{su3constr} and find the realization of the SU(3) transformations
on the physical bosonic variables.

The bosonic core of the involved superfields is as follows
\be
N^{++} = n^{++} + \theta^+\bar\theta^+\,\sigma\,, \quad \omega = \omega_0 + \theta^+ \bar\theta^+\, \sigma^{-2}\,,
\quad q^+ = f^+ + \theta^+\bar\theta^+\,\mu^{-}\,, \; \bar q^+ = \bar f^+ - \theta^+\bar\theta^+\, \bar\mu^-\,, \lb{expan2}
\ee
where all fields in the r.h.s are defined on the manifold $(t, u^\pm_i)$, e.g., $f^+ = f^+(t, u)$, etc.
The basic bosonic constraints are obtained from the superfield ones by the direct replacement
\be
(q^+, \bar q^+, N^{++}, \omega) \;\rightarrow \; (f^+, \bar f^+, n^{++}, \omega_0), \lb{replace}
\ee
The constraints on these fields
literally mimic \p{su3constr}, in which one should just put $\theta = 0$ and make the replacements \p{replace}. It will
be more convenient to start from the equivalent constraints \p{hol1} - \p{om2}, in which case we obtain
\bea
&& (\partial^{++} + \phi^{++})\,f^+ = 0\,, \quad \partial^{++}\phi^{++} + (\phi^{++})^2 = 0\,,\lb{hol1a} \\
&& (\partial^{++}+ \bar\phi^{++})\bar f^+ = 0\,, \quad \partial^{++}\bar\phi^{++} + (\bar\phi^{++})^2 = 0\,, \lb{anthola} \\
&& \phi^{++} - \bar\phi^{++} = -2\alpha_0\,f^+\bar f^+\,, \lb{phiconda} \\
&& \partial^{++}\omega_0 - \phi^{++} -\bar\alpha\,f^+\bar f^+ = 0\,,
\quad \mbox{or} \quad  \partial^{++}\omega_0
- \bar\phi^{++} +\alpha\,f^+\bar f^+ = 0\,, \lb{om2a} 
\eea
where $\phi^{++} = \Phi^{++}_{\theta = 0}\,$.
The $\phi^{++}$-constraints in \p{hol1a} and \p{anthola} can be solved as
\be
\phi^{++} = \frac{A^{++}}{1 + A^{+-}}\,, \qquad \bar\phi^{++} = \frac{\bar A^{++}}{1 + \bar A^{+-}}\,, \quad
\partial^{++}A^{++} = \partial^{++}\bar A^{++} = 0\,, \lb{phisol}
\ee
whence
\be
A^{++} = A^{(ik)}u^+_iu^+_k\,, \;   \bar A^{++} = \bar A^{(ik)}u^+_iu^+_k\,, \quad \phi^{++} = \partial^{++}\ln (1 + A^{+-})\,, \;
\bar\phi^{++} = \partial^{++}\ln (1 + \bar A^{+-})\,. \lb{AbarA}
\ee
Then the constraints for $f^+$ and $\bar f^+$ in  \p{hol1a} and \p{anthola} are solved as
\be
f^{+} = \frac{f^iu^+_i}{1 + A^{+-}}\,, \quad \bar f^{+} = \frac{\bar f^iu^+_i}{1 + \bar A^{+-}}\,.
\ee
The complex field $A^{ik}$ can be divided into  real and imaginary parts
\be
A^{ik} = a^{ik} + i b^{ik}\,.
\ee
The real field $a^{ik}$ is just an analog of the field $a^{ik}$ of the U(2) model and it parametrizes the SU(2) part of the SU(3)  manifold.
The doublet fields $f^k, \bar f^k$, together with the appropriate analog of the singlet field $b$,
complement this SU(2) manifold to the whole SU(3). Using the algebraic constraint \p{phiconda}, $b^{ik}$ can be expressed in terms of $a^{ik}$ and $f^i, \bar f^i$:
\be
b^{ik} = \frac{2\alpha_0}{ 2 + a^2}\Big[\ell^{ik} - \ell^{j(i}a^{k)}_{j} + \frac{1}{2}(\ell\cdot a)a^{ik} \Big], \qquad
\ell^{ik} := if^{(i}\bar f^{k)}\,.
\ee
Its useful corollaries are
\be
b\cdot a = \alpha_0\, \ell\cdot a\,, \quad  b^2 = \alpha_0(\ell\cdot b) =
\frac{2\alpha_0^2}{ 2 + a^2}\Big[\ell^2 + \frac12 (\ell\cdot a)^2\Big].
\ee
The constraint \p{phiconda}, in terms of the harmonic projections, can be also written as
\be
A^{++}(1 + \bar A^{+ -}) - \bar A^{++}(1 + A^{+ -}) = -2\alpha_0\, f^i\bar f^k u^+_i u^+_k = 2i\alpha_0\, \ell^{ik}u^+_i u^+_k\,.\lb{Aconstr1}
\ee
In terms of the ordinary fields $A^{ik}, \bar A^{ik}$, it amounts to
\be
A^{ik} - \bar A^{ik} +A^{(i l}\bar A_l^{\,k)} = -2\alpha_0\, f^{(i}\bar f^{k)}\,.\lb{constr2}
\ee
Note also the relation
\be
A^{++}A^{--} - (A^{+-})^2 = \frac12 A^2
\ee
(and the analogous relation for the complex-conjugated fields). Some useful relations following from the constraint \p{constr2} are
\bea
&& f^{(i}\bar f^{k)}A_{ik} = \frac1{2\alpha_0}(A\cdot\bar A - A^2)\,, \quad f^{(i}\bar f^{k)}\bar A_{ik} =
-\frac1{2\alpha_0}(A\cdot\bar A - \bar A^2)\,, \lb{ffA1} \\
&& f^{(i}\bar f^{k)}A_{k}^m = \frac1{4\alpha_0}\big[ \varepsilon^{im}(A^2 - A\cdot \bar A) + 2\bar A^{k(i}A^{m)}_k +
A^{im}(A\cdot \bar A) -  \bar A^{im} A^2\big], \lb{ffA2} \\
&& f^{(i}\bar f^{k)}\bar{A}_{k}^m = -\frac1{4\alpha_0}\big[ \varepsilon^{im}(\bar{A}^2 - A\cdot \bar A) - 2\bar A^{k(i}A^{m)}_k +
\bar{A}^{im}(A\cdot \bar A) -  A^{im} \bar{A}^2\big], \lb{ffA3} \\
&& (2 +A^2)(2 + {\bar A}^2) = B_+\, B_-\,, \quad B_\pm := 2 + A\cdot \bar A \pm 2\alpha_0\,\bar f^if_i\,, \lb{BpmDef} \\
&&(2 + A\cdot \bar A)^2 - (2 +A^2)(2 + {\bar A}^2) = 4\alpha_0^2 (\bar f^if_i)^2\,, \lb{Norm} \\
&& (\bar f^i f_i)\partial_t{({\bar f}^i f_i)} = \frac1{4\alpha_0^2}\big[\partial_t({A}\cdot \bar A)(2 + A\cdot \bar A)-
(\dot A\cdot A) (2 + \bar A^2) -(\dot{\bar A}\cdot \bar A)(2 + \bar A^2)\big]. \lb{lastId}
\eea

Using the superfield transformation laws \p{Phitr} and \p{Phiu2},  we can find the SU(3)/U(2) transformations
of the ``central basis'' fields $A^{ik}$ and $f^i, \bar f^i$:
\bea
&& \delta A^{ik} = \alpha_0\Big[(\bar\xi_lf^l)A^{ik} - 2\bar\xi^{(i} f^{k)}\Big], \quad
\delta \bar A^{ik} = \alpha_0\Big[(\xi^l\bar f_l)\bar A^{ik} + 2\xi^{(i} \bar f^{k)}\Big], \lb{Atr} \\
&& \delta f^k = \xi^k - A^{kl}\xi_l + \alpha_0 (\bar\xi_lf^l)f^k\,, \quad
\delta \bar f^k = \bar\xi^k - \bar A^{kl}\bar\xi_l + \alpha_0 (\xi^l\bar f_l)\bar f^k\,, \lb{ftr}
\eea
as well as their U(2) transformations
\bea
\delta A^{ik} &=& 2\alpha_0\,[\lambda^{ik} + \frac{1}{2}(\lambda\cdot A)A^{ik} +\lambda^{(il}A_l^{k)}], \;
\delta \bar A^{ik} = 2\alpha_0\,[\lambda^{ik} + \frac{1}{2}(\lambda\cdot \bar A)\bar A^{ik} +\lambda^{(il}\bar A_l^{k)}], \lb{Au2} \\
\delta f^i &=& \alpha_0\,[ 3i\phi f^i + (\lambda\cdot A)f^i], \quad \delta \bar f^i = \alpha_0\,[ -3i\phi \bar f^i
+ (\lambda\cdot \bar A)\bar f^i]. \lb{fu2}
\eea

An interesting point is that, like in the previously considered U(2) case,  there exists another SU(2) which commutes with the
SU(3) transformations given above. It is realized by the transformations
\bea
\delta A^{ik} &=& 2\alpha_0\,[-\tau^{ik} - \frac{1}{2}(\tau\cdot A)A^{ik} +\tau^{(il}A_l^{k)}], \nn
\delta \bar A^{ik} &=& 2\alpha_0\,[-\tau^{ik} - \frac{1}{2}(\tau\cdot \bar A)\bar A^{ik} +\tau^{(il}\bar A_l^{k)}], \lb{Au22} \\
\delta f^i &=& \alpha_0\,[- (\tau \cdot A)f^i + 2\tau^{(il)} f_l], \quad \delta \bar f^i = \alpha_0\,[-(\tau\cdot \bar A)\bar f^i
 + 2\tau^{(il)} \bar{f}_l]. \lb{fu22}
 \eea
They have the same closure as the SU(2) transformations  \p{Au2}, \p{fu2}  and the analogous ones considered in subsection 2.2. The diagonal
SU(2) in the product of \p{Au2}, \p{fu2} and \p{Au22}, \p{fu22}  is just the harmonic group SU(2)$_H$ (with the parameters rescaled by $2\alpha_0$).
It is easy to check the covariance
of the constraint \p{constr2} under these transformations and to be convinced that the superfield constraints \p{hol1} - \p{phicond} are
also covariant. This last property immediately follows from the fact that the generators of the extra SU(2) are proper linear combinations
of those generating the SU(2) part of the transformations \p{Phiu2} and of the generators of the harmonic SU(2). By the same reasoning,
the action \p{actsup2} is invariant. Moreover, we can extend the extra SU(2) to U(2) by noting that the superfield constraints
and the action \p{actsup2} are invariant under an extra U(1) acting as constant shifts of the superfield $\omega$ and its first bosonic component $\omega_0$:
\be
\omega' = \omega + \tau_0 \quad \Leftrightarrow \quad \omega_0' = \omega_0 + \tau_0\,.\lb{tau0}
\ee
 So we come to the conclusion that the invariance group of the system we are considering
is the product SU(3)$\times$U(2). Only SU(3)$\times$U(1) in this product commutes with ${\cal N}=4$ supersymmetry
and so defines the triholomorphic isometries \footnote{We inquired whether one can extend \p{Au22}, \p{fu22} to another (right) SU(3)
 which would (a) commute with the first (left) SU(3) and (b) preserve the important constraint \p{constr2}. We failed to find such an extension.}.

In analogy with the SU(2)$\times$U(1) case, the corresponding bosonic nonlinear sigma model should be associated with the eight-dimensional coset manifold
[SU(3)$_L\times$U(2)$_R]/$U(2)$_{diag}\,$, where U(2)$_{diag}$  involves SU(2)$_H$\footnote{The U(1) factor in U(2)$_{diag}$ acts as homogeneous phase
transformations of $f^i$ and $\bar{f}^i$ (or of the superfields $q^{+}, \bar{q}^+$, in the original setting).}.  If $g(\varphi)$ is an element of SU(3) and $\varphi^A, A=1, \ldots, 8,$ are
local coordinates on SU(3), the nonlinear sigma model Lagrangian can be constructed from the $d=1$ pullbacks of the current
\be
{\cal J}_A = g^{-1}\partial_A g\,. \lb{Current}
\ee
A metric on SU(3) respecting at least SU(3)$_L$ invariance then reads
\be
g_{AB} = {\rm tr}({\cal J}_A \,X \,{\cal J}_B)\,, \lb{MetRic}
\ee
where $X$ is a $3\times 3$ matrix which should be chosen as $X = {\rm diag}(\kappa, \kappa, \chi)$ in order to preserve the SU(3)$_L\times$U(2)$_R$ symmetry. Here $\kappa$ and
$\chi$ are some numerical parameters. At
$\kappa = \chi$, {\it i.e.} for $X = \kappa\,\mathbb{I}{}_{(3\times 3)}\,$, the symmetry is enhanced  to the product SU(3)$_L\times$SU(3)$_R$, while for $\kappa \neq \chi$ SU(3)$_R$ is broken to U(2)$_R$.
Just the latter option is expected to be valid in our ${\cal N}=4$ SU(3) model, with the ratio of the parameters $\chi$ and $\kappa$ strictly fixed.
In order to check all this, we need the explicit expression for the bosonic invariant
action as an integral over $t$. This form of the bosonic action (with the harmonic integrals explicitly done) will be presented elsewhere.

\subsection{Solving further constraints}

Our  ultimate purpose is to find the bosonic component Lagrangian. Besides the constraints on the fields
$\omega_0, f^i, \bar f^i$ and $n^{++}$,
we now need also those for the remaining bosonic components in the $\theta$ expansions \p{expan2}, because all such components
are involved in the bosonic action. The harmonic equations for the extra fields $\sigma, \omega_0, \sigma^{-2}$ and $\mu^-, \bar\mu^-$ following
from the superfield constraints read
\bea
&&(\partial^{++} + 2\phi^{++})\tilde{\sigma} - 2i\dot{\phi}{}^{++} = 0\,, \quad
(\partial^{++} + 2\bar\phi^{++})\bar{\tilde{\sigma}} + 2i\dot{\bar\phi}{}^{++} = 0\,, \lb{sigma} \\
&& \partial^{++}\omega_0 - \frac{1}{2}\beta_{(+)}\,\phi^{++}  - \frac{1}{2}\beta_{(-)}\,\bar\phi^{++} = 0\,, \lb{omegaeq} \\
&& \partial^{++}\sigma^{-2} -2i\dot{\omega}_0 - \frac{1}{2}\left(\tilde{\sigma}-\bar{\tilde{\sigma}}\right) +
i\alpha_1(\mu^-\bar{f}{}^+ -\bar\mu{}^- f^+) = 0\,, \lb{sigma-2} \\
&&(\partial^{++} + \phi^{++})\mu^- + \tilde{\sigma}f^+ - 2i\dot{f}{}^+=0\,, \quad (\partial^{++} + \bar\phi^{++})\bar\mu^- + \bar{\tilde{\sigma}}\bar f^+
+ 2i\dot{\bar f}{}^+ = 0\,,\lb{mueq}
\eea
where
\be
\tilde{\sigma} = \sigma - \alpha\,(\mu^-\bar{f}{}^+ -\bar\mu{}^- f^+)\,, \quad \bar{\tilde{\sigma}} = -\sigma -
\bar\alpha\,(\mu^-\bar{f}{}^+ -\bar\mu{}^- f^+)\,, \quad \beta_{(\pm)} = 1 \pm i\,\frac{\alpha_1}{\alpha_0}\,,\lb{tildedef}
\ee
and we took advantage of the constraint \p{phiconda}.

Using \p{AbarA}, it is easy to solve eq. \p{omegaeq}:
\be
\omega_0 = b + \frac{1}{2}\beta_{(+)} \ln (1 + A^{+-}) + \frac{1}{2}\beta_{(-)} \ln (1 + \bar A^{+-})\,,\lb{omega0}
\ee
where $b = b(t)$ is a real harmonic-independent $d=1$ field. Also, the general solution of eqs. \p{sigma} is
\be
\tilde{\sigma} = \frac{1}{(1 + A^{+-})^2}\Big\{c + 2i\,[\dot{A}{}^{+-} - A^{il}\dot{A}^k_l u^+_{(i}u^-_{k)}]\Big\}\,, \mbox{and\; c.c.}\,,\lb{sigmaexpr}
\ee
where $c = c(t)$ is a complex harmonic-independent $d=1$ field. Note that \p{sigmaexpr} and the definition \p{tildedef} allow one to find
\be
\mu^-\bar{f}{}^+ -\bar\mu{}^- f^+ = -\frac{1}{2\alpha_0}\left(\tilde{\sigma} + \bar{\tilde{\sigma}} \right).\lb{muf}
\ee
Further, taking the harmonic integral of eq. \p{sigma-2} and making use of
the formulas from Appendix, we find one relation between the fields $c$ and $b$:
\be
\beta_{(+)}\,\frac{1}{2 + {A^2}}\,(c + i \dot{A}\cdot A) - \beta_{(-)}\,\frac{1}{2 + {\bar A{}^2}}\,
(\bar c - i \dot{\bar A}\cdot \bar A) + 2i\dot{b} = 0\,. \lb{c-b}
\ee

Like in the case of the $SU(2)\times U(1)$ model, to compute the component action there is no need to explicitly solve eq. \p{sigma-2}.
However, one still needs to have the solution of eqs. \p{mueq}. It is convenient to redefine
\be
\mu^- = \frac{1}{1 + A^{+-}}\,\hat{\mu}{}^{-}\,, \quad \partial^{++}\hat{\mu}{}^- - 2i\,\dot{f}^iu^+_i + f^iu^+_i\left(\tilde{\sigma}
+ 2i\frac{\dot{A}^{+-}}{1 +  A^{+-}}\right) = 0\,.
\ee
After some work, the solution is found to be
\bea
&&\hat{\mu}^- = 2i\dot{f}^iu^-_i + \omega^-\,c(t) - 2i\dot{A}^{(jk)}\left(f^iu^-_i\, A_{(jk)}\,\frac{1}{2 + A^2} + f^iu^+_i\,\Psi^{--}_{(jk)}\right), \label{hatmu1} \\
&&\hat{\bar{\mu}}^- = -2i\dot{\bar{f}}^iu^-_i + \bar\omega^-\,\bar c(t) + 2i\dot{\bar{A}}^{(jk)}\left(\bar{f}^iu^-_i\, \bar{A}_{(jk)}\,\frac{1}{2 + \bar{A}^2} + \bar{f}^iu^+_i\,\bar\Psi^{--}_{(jk)}\right), \label{hatmu2}
\eea
where
\bea
\omega^- = - \frac{2}{(2 + A^2)}\left( f^ku^-_k\ - f^k u^+_k\, \frac{A^{--}}{1 + A^{+-}}\right), \;
\bar\omega^- = - \frac{2}{(2 + \bar{A}^2)}\left( \bar{f}^ku^-_k\ - \bar{f}^k u^+_k\, \frac{\bar{A}^{--}}{1 + \bar{A}^{+-}}\right), \lb{omlam}
\eea
\bea
\Psi^{--}_{ik} = \frac{1}{(2 + A^2)(1 + A^{+-})}\Big\{u^-_iu^-_k\,[2 + \frac{1}{2}A^2 - (A^{+-})^2]+ [2u^+_{(i}u^-_{k)}A^{+-} -
u^+_iu^+_k A^{--}]A^{--}\Big\}.  \lb{psisol}
\eea
Note the useful formula
\bea
\dot{A}^{ik}\Psi^{--}_{ik} = \frac{1}{(2 + A^2)(1 + A^{+-})}\Big[\dot{A}^{--}(2 + A^2) - (\dot A\cdot A)A^{--}\Big].
\eea

Now, using the explicit expressions for $\mu^-, \bar\mu^-$ given above and the relation \p{muf}, we can establish one more relation between
fields $c(t)$ and $b(t)$, in addition to \p{c-b}
\be
\frac{1}{2 + {A^2}}\,(c + i \dot{A}\cdot A) + \frac{1}{2 + {\bar A{}^2}}\,
(\bar c - i \dot{\bar A}\cdot \bar A) = 2i\alpha_0\,D\,, \lb{c-b1}
\ee
where
\be
D := \frac{1}{2 + (A\cdot \bar A) - 2\alpha_0\,\bar f^if_i}\Big[\dot f^i\bar f_i - f^i\dot{\bar f}_i +
\frac{1}{2\alpha_0}(\dot A\cdot \bar A - \dot{\bar A}\cdot A)  \Big], \quad \bar D = -D\,.
\ee
The set of eqs. \p{c-b} and \p{c-b1} can be solved for $c$ and $\bar c$ as
\be
c + i(\dot A\cdot A) = -i(2 + A^2)\Big[ \dot b - \alpha_0\,\beta_{(-)}\,D\Big], \quad \bar c - i(\dot{\bar A}\cdot \bar A) =
i(2 + \bar A^2)\Big[ \dot b + \alpha_0\,\beta_{(+)}\, D\Big].
\ee
In what follows, we  will also use
\be
\hat c := c(b=0) = -i \Big[ (\dot A\cdot A) - \alpha_0(2 + A^2)\beta_{(-)}\,D\Big], \;
\hat{\bar c} =i \Big[ (\dot{\bar A}\cdot \bar A) + \alpha_0(2 + \bar A^2)\beta_{(+)}\,D\Big].
\ee

To close this subsection, we present the SU(3) transformation laws of the field $b(t)$. They can be found by comparing the $\theta=0$
part of the transformations \p{omsu3}, \p{omu2} with the direct variation of $\omega_0$ in \p{omega0}:
\be
\delta_{(\xi)} b = -\frac{\alpha_0}{2}\Big[\beta_{(-)}(\xi^l\bar f_l) + \beta_{(+)}(\bar\xi_if^i)\Big], \quad
\delta_{(\phi,\lambda)} b = {\alpha_1}\, \phi  -
\frac{\alpha_0}{2}\Big[\beta_{(+)}(\lambda\cdot A) + \beta_{(-)}(\lambda\cdot\bar A)\Big]. \lb{Transf_b}
\ee

It is easy to construct the covariant derivative ${\cal D}_t b$. We first notice that under the $\xi_i$ and $\lambda$ transformations:
\bea
&&\delta_{(\xi)} B_- = \alpha_0 (\xi^i {\bar f}_i +  \bar\xi_i {f}^i)\, B_-\,, \quad  \delta_{(\xi)} D =
-\frac12( \xi^i\dot{\bar f}_i - \bar\xi_i\dot{f}^i), \lb{TranBD1} \\
&& \delta_{(\lambda)} B_-  = \alpha_0 \big(\lambda\cdot A +  \lambda\cdot \bar A\big)\,, \quad  \delta_{(\lambda)} D =\frac12\big(\lambda\cdot \dot{A} - \lambda\cdot \dot{\bar A}\big)\,,
\eea
where $B_-$ was defined in \p{BpmDef}. Then
\be
{\cal D}_t b = \dot{b} + i\alpha_1 D + \frac12 B^{-1}_-\dot{B}_-\,, \quad \delta_{(\xi)} {\cal D}_t b = \delta_{(\phi,\lambda)} {\cal D}_t b =0\,.\lb{CovDp}
\ee
{}From this definition it follows:
\be
({\cal D}_t b)^2 = (\dot b)^2 + 2i\alpha_1\dot b D + \dot b B^{-1}_-\dot{B}_- - \alpha_1^2 D^2 + \frac14 B^{-2}_-(\dot B_-)^2 +
i\alpha_1D B^{-1}_-\dot{B}_-\,.\lb{squareDtb}
\ee
Like in the $U(2)$ case, the covariant derivative ${\cal D}_t b$ is simplified after the redefinition
\be
\tilde{b} = b+ \frac12 \ln B_-\,, \quad  {\cal D}_t b = \dot{\tilde{b}} + i\alpha_1 D\,.
\ee
Finally, we note that the extra U(1) symmetry \p{tau0} acts as a constant shift of the field $b$ (or of $\tilde{b}$):
\be
b' = b + \tau_0\,.
\ee
The group U(1)$\,\subset\,$SU(3) also acts as a shift of $b$, but it acts as well on $f^i, \bar f_i$ as a phase
transformation. The extra U(1) affects only the field $b(t)$ \footnote{The homogeneous part of the product of these two $U(1)$ transformations belongs to
the stability subgroup U(2)$_{diag}$ of the coset [SU(3)$_L\times$ U(2)$_R]/$U(2)$_{diag}$, while the remainder amounts to a shift of $b$ as the 8-th coset parameter.}.
It is also worth noting that under the extra SU(2) symmetry \p{Au22}, \p{fu22}
the field $b$ transforms as
$$
\delta_{(\tau)} b = \frac{\alpha_0}{2}\Big[\beta_{(+)}(\tau\cdot A) + \beta_{(-)}(\tau\cdot\bar A)\Big].
$$
The covariant derivative ${\cal D}_t b$ is invariant under the $\tau$ transformations too, $\delta_{(\tau)} {\cal D}_t b =0\,$.

\subsection{The bosonic invariant action}
The invariant action \p{actsup2} consists of two parts, which can be written as
\be
S_{su(3)} = S(N,\omega) + \alpha_0\,S(q).\lb{actsup3a}
\ee
After integrating over Grassmann coordinates and passing to the bosonic limit, we obtain
\bea
S(N,\omega) \rightarrow  S_1 = \frac{i}{2}\int dt du \left(n^{++}\dot{\sigma}^{-2} + \sigma\,\dot\omega_0 \right), \quad
S(q)  \rightarrow  S_2 = i\int dt du \left(\bar\mu{}^-\dot{f}{}^+ + \mu^- \dot{\bar{f}}{}^+\right),\lb{compact22}
\eea
and so
\be
S_{su(3)}^{bos} = S_1 + \alpha_0\,S_2.\lb{actsup3}
\ee
In Appendix A we will present the proof that \p{actsup3}
is indeed SU(3) invariant like its full superfield prototype \p{actsup2}.

To compute the bosonic  action as a $t$-integral, one needs to substitute the explicit expressions of the harmonic fields that were
found in the previous subsection and then do the harmonic integrals. The latter task is rather difficult, because,
as opposed to the U(2) case, we face here integrals involving in the denominator products of the harmonic
factors $( 1 + A^{+-})$ and $(1 + \bar{A}^{+-})$ with $A^{ik} \neq \bar A^{ik}$. Some basic integrals of this kind are calculated in Appendix B.
We postpone the purely technical task of restoring the full bosonic component action to the next
publication. Here we limit our consideration only to the $b$-dependent part of the  bosonic action \p{actsup3}.

To find this part,  we should collect those terms in \p{actsup3} which involve the field $b$. The full set of such terms is as follows
\bea
S(b) &=& \int dt \Big\{2 \dot{b}\Big(\frac{\dot A\cdot A}{2 + A^2} + \frac{\dot{\bar A}\cdot \bar A}{2 + \bar A^2} \Big)
 - \frac12\dot{b}\Big[ (2 + A^2)\int du \frac{\dot{\bar A}^{+-}}{(1 + A^{+-})^2(1 + \bar A^{+-})} \nn
&& +\, (2 + \bar A^2)\int du \frac{\dot{A}^{+-}}{(1 + \bar A^{+-})^2(1 + A^{+-})}\Big] + 2i\alpha_1\,\dot b\,D + (\dot b)^2 \nn
&& -\, \frac12 \ddot{b}\left( 2 + A\cdot \bar A - 2\alpha_0\,\bar f^if_i\right) \int du \frac{1}{(1 + A^{+-})(1 + \bar A^{+-})}\Big\}\,.\lb{b-terms}
\eea
Here, the first two lines are the contribution from $S_1$, while the third line comes from $S_2$.
The explicit expressions for the relevant harmonic integrals are collected in Appendix B. Using them and integrating by parts in the
last line of \p{b-terms}, we finally obtain
\be
S(b) = (\dot b)^2 + 2i\alpha_1\dot b D + \dot b B^{-1}_-\dot{B}_-\,, \lb{b-part}
\ee
in accordance with the formula \p{squareDtb}. The $b$-independent terms completing $S(b)$ to the total expression \p{squareDtb} should
come from the remainder of the bosonic action \p{actsup3}.

\setcounter{equation}{0}

\section{Summary and outlook}
The basic aim of the present paper was to start the construction of the harmonic superfield actions for the group manifolds
with quaternionic structure as nice explicit examples of the ``strong'' HKT ${\cal N}=4$ supersymmetric  $d=1$ sigma models
the general formulation of which was given in \cite{DIN4}. We limited our attention to the simple cases of the 4-dimensional group manifold U(2)
and the 8-dimensional one SU(3). For both cases we have found the relevant nonlinear harmonic constraints and shown that the relevant invariant actions
are bilinear in the superfields involved, as it should be for the strong HKT models in agreement with the reasoning of \cite{DIN4}. It was found that
the full internal symmetry of the U(2) and SU(3) models are, respectively, U(2)$\times$SU(2) and SU(3)$\times$U(2), with the standard harmonic SU(2)$_H$
symmetry forming the diagonals in these products. For the U(2) case we computed the full bosonic action, which is none other than  the $S^3\times S^1$ one,
and showed that the gauging of its U(1) isometry, both at superfield and at component levels, yields a particular case of the nonlinear $({\bf 3, 4, 1})$
multiplet action, with pure $S^3$ bosonic target. For the SU(3) case we presented the bosonic action in the space $(t, u^\pm_i)$, independently checked its
SU(3)$\times$U(2) invariance and gave explicitly an important part of it  by doing the integration over harmonic variables. For the U(2) case we also performed
the detailed comparison with the general harmonic approach to HKT geometries (of refs. \cite{DIN4} and \cite{FIS}) and found excellent agreement
with this approach. We explicitly constructed the closed torsion for this case, as well as the Bismut and Obata connections, with respect to which the corresponding
triplet of quaternionic complex structures is covariantly constant. The Riemann curvatures of both these connections were checked to vanish.

It may be possible to generalize the considerations of section 2 to bigger groups. One could consider harmonic superfields
$N^{++}$ and $\Omega$ which are $n$ by $n$ matrices, satisfying the constraints
\begin{equation}\begin{array}{l}D^{++}\Omega +N^{++}\Omega=0,\cr D^{++}N^{++} +(N^{++})^2=0.\end{array}
\end{equation}
These constraints are covariant under the following transformation laws
\begin{equation}\begin{array}{l}\delta\Omega=(-\Lambda^{+-}+\rho+N^{++}\Lambda^{--})\Omega,\cr
\delta N^{++}=\Lambda^{++} -\{\Lambda^{+-},N^{++}\}+[\rho,N^{++}]+N^{++}\Lambda^{--}N^{++},\label{transm}
\end{array}\end{equation}
where $\rho$ and $\Lambda^{\pm\pm}$ are $n\times n$ matrices. Elements of $\rho$ are constants, and elements of $\Lambda^{\pm\pm}$ are triplets of the harmonic
$SU(2)$, $\Lambda^{\pm\pm}=\Lambda^{ij}u^\pm_iu^\pm_j$.
These transformations form an algebra
\begin{equation}\begin{array}{l}
[\delta,\delta'](\Omega, N^{++})=\delta^"(\Omega, N^{++}),\cr
\rho"=-[\rho,\rho']-[\Lambda^{+-},\Lambda'^{+-}]+\frac{1}{2}[\Lambda^{++},\Lambda'^{--}]+\frac{1}{2}[\Lambda^{--},\Lambda'^{++}],\cr
\Lambda"^{+-}= -\frac{1}{2}\{\Lambda^{++},\Lambda'^{--}\}+ \frac{1}{2}\{\Lambda^{--},\Lambda'^{++}\}+[\rho',\Lambda^{+-}]-[\rho,\Lambda'^{+-}]
\end{array}.\end{equation}
The bosonic part of the superfields $N^{++}$ and $\Omega$ should parametrize the group SU($2n$)$\times$U(1), and the transformations in (\ref{transm})
correspond to infinitesimal SU($2n$)$\times$U(1) translations. These fields generalize those in section 2. However, we have not been able
as yet to write an invariant action generalizing (\ref{actNom}). We leave this problem for further study.

It remains as well to explicitly derive the complete component action for the SU(3) model, and to recover the basic geometric objects of this model from the
general harmonic HKT formalism of refs. \cite{DIN4} and \cite{FIS}, as it was done for the U(2) model in section 3.

\setcounter{equation}{0}
\section*{Acknowledgments}
The authors thank Andrei Smilga for his interest in this work and valuable discussions. The work of E.I. was supported
by the RFBR grant, project No 18-02-01046, and a grant of the Heisenberg-Landau program.
He thanks the Laboratoire de Physique, ENS-Lyon, for the kind hospitality extended to him a few times in the course of this
work. A part of this research was completed, when E.I. was visiting SUBATECH, University of Nantes.
He is indebted to the Directorate of SUBATECH for its kind hospitality.

\bigskip
\renewcommand\theequation{A.\arabic{equation}} \setcounter{equation}0
\section*{Appendix A\quad  }

In this Appendix we prove the SU(3) invariance of  the bosonic action \p{actsup3}. The proof follows the same line as the check of the invariance
of the superfield action \p{actsup2}: we consider only
the coset SU(3)/U(2) transformations and leave in them only holomorphic parts $\sim \xi_i$. We do not assume in advance the relations \p{specalpha}, \p{specalpha1}
except for $\gamma_1  = -\alpha_1$. Then the component field transformations we are interested in read
\bea
&&\delta f^+ = \xi^+ - \xi^-\phi^{++}\,, \qquad \delta \bar f^+ = -2\alpha_0\, \xi^-(\bar f^+)^2\,, \nn
&& \delta n^{++} = \alpha\big(\xi^+ - \xi^- n^{++} \big)\,\bar{f}^+ - \alpha\bar\alpha \,\xi^- f^+(\bar f^+)^2\,, \nn
&& \delta \phi^{++} = 0\,, \qquad \delta \bar\phi^{++} = 2\alpha_0 \big(\xi^+  - \xi^-\bar\phi^{++}\big)\,\bar{f}^+\,, \nn
&& \delta \sigma = -\alpha\, \xi^+\bar\mu^- +\alpha\, \xi^- \big(n^{++} \bar\mu^- - \sigma \bar{f}^+\big) - \alpha \bar\alpha\,\xi^- \big[(\bar f^+)^2\mu^- - 2f^+\bar{f}^+ \bar\mu^-\big]\,, \nn
&&\delta \omega_0 = \bar\alpha\, \xi^- \bar{f}^+\,, \qquad \delta \sigma^{-2} = -\bar\alpha\, \xi^-\bar\mu^-\,, \nn
&& \delta \mu^- = -\xi^-\big[\sigma + \alpha \big(f^+ \bar\mu^- - \bar{f}^+ \mu^- \big)\big], \qquad \delta \bar{\mu}^- = -4\alpha_0\, \xi^- \bar{f}^+\bar\mu^-\,. \lb{compTran}
 \eea
These are directly deduced by substituting the truncated superfields \p{expan2} into the transformation laws \p{su3u2}, \p{omsu3} and \p{Phitr}. Substituting
$\xi^+ = \partial^{++}\xi^-\,,$ integrating by parts with respect to $\partial^{++}$ and $\partial_t$, using the constraint \p{omegaeq} alongside with the second constraint in \p{mueq},
as well as the relations \p{tildedef} and \p{muf},
we are able to cast the variation $\delta S_1$ in the form
\bea
\delta S_1 &=& \frac{i}{2}\int dt du \,\xi^-\Big\{ \frac{\alpha}{2}\,\big[\beta_{(+)}\tilde{\sigma} -  \beta_{(-)}\bar{\tilde{\sigma}}\big] + \bar\alpha\,\big(\dot{n}^{++} \bar{\mu}^-  +
\sigma \dot{\bar{f}}^+\big) \nn
&& -\, \alpha\,\dot{\bar\mu}^- \big(n^{++} - 2i \alpha_1\,f^+\bar{f}^+\big) \Big\}. \lb{VarS1}
\eea
Analogously,
\bea
\delta S_2 &=& {i} \int dt du\, \xi^- \Big\{-4\alpha_0 {\bar\mu}^- \bar{f}^+ \dot{f}^+  + \dot{\bar\mu}^-\phi^{++} - \sigma \dot{\bar{f}}^+ \nn
&& -\, \alpha \big(f^+ {\bar\mu}^-
- \bar{f}^+ {\mu}^- \big)\dot{\bar{f}}^+ + 2\alpha_0 \dot{\mu}^- (\bar{f}^+)^2 \Big\}.      \lb{VarS2}
\eea
For $\delta S^{bos}_{su(3)} = \delta S_1 + \alpha_0 \delta S_2$ we obtain, up to total $t$-derivative and making use of \p{muf},
\bea
\delta S^{bos}_{su(3)} &=& i \int dt du\,\xi^-\Big\{ \dot{f}^+ \bar{f}^+\bar{\mu}^- \big( \alpha\bar\alpha\ - 4\alpha_0^2\big) + \dot{\bar{f}}^+f^+ \bar{\mu}^-\big[ i\alpha_1\alpha_0 \beta_{(+)}
+ \alpha\bar\alpha - \alpha\alpha_0\big] \nn
&&+\,  \dot{\bar{f}}^+\bar{f}^+{\mu}^-\big[ \alpha\alpha_0 - 4\alpha_0^2 -i\alpha_1\alpha_0 \beta_{(+)}\big] \Big\}. \lb{VarS}
\eea
All terms in \p{VarS} are independent, so in order to  have $\delta S^{bos}_{su(3)} =0$, their coefficients should  vanish separately,
\bea
({\rm a})\,\alpha\bar\alpha\ - 4\alpha_0^2 = 0\,, \quad ({\rm b})\,i\alpha_1\alpha_0 \beta_{(+)} + \alpha\bar\alpha - \alpha\alpha_0 =0\,, \quad
({\rm c})\,\alpha\alpha_0 - 4\alpha_0^2 -i\alpha_1\alpha_0 \beta_{(+)} =0\,. \lb{CondCompo}
\eea
It is direct to check that the conditions (\ref{CondCompo}b) and (\ref{CondCompo}c) are identically satisfied as a consequence of the
single condition (\ref{CondCompo}a),
\be
\alpha\bar\alpha\ - 4\alpha_0^2 = 0 \quad \Longleftrightarrow \quad \alpha_1^2 = 3\alpha_0^2\,,\lb{BasicCondo}
\ee
which is just the relation already found in \p{specalpha1}.

Thus we have independently proved that the bosonic action \p{actsup3} is SU(3) invariant under the condition \p{BasicCondo}
and so is guaranteed to define some $d=1$ sigma model on the SU(3) group manifold. It is also
straightforward to show its invariance under the transformations of the U(2) group commuting with SU(3). It is worth pointing out that checking all these
invariances does not require doing explicitly the harmonic integrations in \p{actsup3}.

\renewcommand\theequation{B.\arabic{equation}} \setcounter{equation}0
\section*{Appendix B\quad  }
In this Appendix we present the calculation of some harmonic integrals. The general method consists in expanding the harmonic integrands
in power series in $a^{+-} = a^{(ik)}u^+_{(i}u^-_{k)}$ and then doing the harmonic integrals for each term in this expansion using
the general formula
\be
\int du (a^{+-})^{2n} = (-1)^n \frac{1}{2n + 1}\left(\frac{a^2}{2}\right)^n\,, \quad a^2 = a^{(ik)}a_{(ik)}\,.
\ee
The integral of any odd power of $a^{+-}$ is vanishing. In this way, we obtain, e.g.,
\bea
&&\int du \,\ln (1 + a^{+-}) = -1 +\frac{1}{2}\,\ln \left(1 +\frac{a^2}{2}\right) + \frac{\arctan \sqrt{\frac{a^2}{2}}}{\sqrt{\frac{a^2}{2}}}\,,
\lb{IntHarm1}\\
&&\int du \,\frac{1}{1 + a^{+-}} = \frac{\arctan \sqrt{\frac{a^2}{2}}}{\sqrt{\frac{a^2}{2}}}\,, \lb{IntHarm2}\\
&&\int du \, \frac{a^{+-}}{(1 + a^{+-})^2} = -\left(\partial_\beta\,\int du\,\frac{1}{1 + \beta a^{+-}}\right)_{\beta =1} =
\frac{\arctan \sqrt{\frac{a^2}{2}}}{\sqrt{\frac{a^2}{2}}} - \frac{1}{1 +\frac{a^2}{2}}\,, \lb{IntHarm3}
\eea
\bea
&&\int du \, \frac{1}{(1 + a^{+-})^2} = \int du\, \left[\frac{1}{1 + a^{+-}} -  \frac{a^{+-}}{(1 + a^{+-})^2} \right] =
\frac{1}{1 +\frac{a^2}{2}}\,,\lb{IntHarm4}\\
&&\int du \, \frac{a^{+-}}{(1 + a^{+-})^3} = -\frac{1}{2}\left(\partial_\beta\,\int du\,\frac{1}{(1 + \beta a^{+-})^2}\right)_{\beta =1}
= \frac{a^2}{2}\,\frac{1}{\left(1 +  \frac{a^2}{2}\right)^2}\,, \lb{IntHarm6}\\
&&\int du \, \frac{1}{(1 + a^{+-})^3} =\int du \, \left[\frac{1}{(1 + a^{+-})^2} - \frac{a^{+-}}{(1 + a^{+-})^3}\right] =
\frac{1}{\left(1 +\frac{a^2}{2}\right)^2}\,.\lb{IntHarm5}
\eea

It is also easy to calculate the integrals of the type
\be
\mbox{(a)} \;I_{(ik)} = \int du\, \frac{u^+_{(i}u^-_{k)}}{(1 + a^{+-})^2}\,, \quad \mbox{(b)} \;I_{(ik)}'
= \int du\, \frac{u^+_{(i}u^-_{k)}}{1 + a^{+-}}\,.\lb{Iikdef}
\ee
Keeping in mind that SU(2) acting on the doublet indices cannot be broken in the process of harmonic integration, these integrals
should be of the form
\be
I_{(ik)} = a_{(ik)}\,f(a^2)\,, \quad I_{(ik)}' = a_{(ik)}\,f'(a^2)\,.\lb{Iik}
\ee
Contracting $I_{(ik)}$ with $a^{(ik)}$ and  using \p{IntHarm3}, we find
\be
I_{(ik)} = a_{(ik)}\,\frac{1}{a^2}\left(\frac{\arctan \sqrt{\frac{a^2}{2}}}{\sqrt{\frac{a^2}{2}}} - \frac{1}{1 +\frac{a^2}{2}} \right).\lb{Iikfin}
\ee
This expression is obviously non-singular at $a^2 = 0\,$. Using \p{Iik}, it is easy to show, e.g., that $a^{il}\dot{a}^k_l I_{(ik)}=0\,$,
which just means  vanishing of the expression \p{Int1} and of the first term in the square bracket in \p{bosAct2}. The explicit form of
the integral $I_{(ik)}'$ can also be easily found
\be
I_{(ik)}' = a_{(ik)}\,\frac{1}{a^2}\left(1 -\frac{\arctan \sqrt{\frac{a^2}{2}}}{\sqrt{\frac{a^2}{2}}}\right).\lb{Iikfinprime}
\ee
It is non-singular at $a^2 = 0\,$, like $I_{(ik)}$.

Using similar reasonings, one can compute
\be
J_{(ik)} = \int du \frac{u^+_{(i}u^-_{k)}}{(1 + a^{+-})^3} = \frac{1}{2}a_{(ik)}\,\frac{1}{\left(1 +\frac{a^2}{2} \right)^2} \lb{J}
\ee
and
\bea
J_{(ik)(mn)} = \int du\,u^+_{(i}u^-_{k)}\,u^+_{(m}u^-_{n)}\,\frac{3 + a^{+-}}{(1 + a^{+-})^3} = a_{(ik)}a_{(mn)} g(a^2) +
(\varepsilon_{im}\varepsilon_{kn} +\varepsilon_{km}\varepsilon_{in}) \tilde{g}(a^2), \lb{J1}
\eea
where
\bea
g(a^2) = \frac{1}{2\left(1 +\frac{a^2}{2} \right)^2}\,, \qquad  \tilde{g}(a^2) = -\frac{1}{4\left(1 +\frac{a^2}{2} \right)}\,.\lb{J2}
\eea

Let us also give an example of splitting some function of $a^{+-}$ into harmonic independent and harmonic dependent parts.
As such function we choose $1/(1 + a^{+-})$ and write
\be
\frac{1}{1 + a^{+-}} = \lambda(a^2) + \partial^{++}\psi^{--}(a^{+-}, a^2). \lb{Ex1}
\ee
Our task is to find $\lambda(a^2)$ and $\psi^{--}(a^{+-}, a^2)$.

Integrating both sides of \p{Ex1} over harmonics and using \p{IntHarm2}, we find
\be
\lambda(a^2) =  \frac{\arctan \sqrt{\frac{a^2}{2}}}{\sqrt{\frac{a^2}{2}}}\,.
\ee
Then we represent $\psi^{--}$ as
\be
\psi^{--} = a^{--}f(a^2, a^{+-}),
\ee
and substitute it into \p{Ex1}, denoting $a^{+-} :=z$. We find
\be
\lambda + [(z^2 + \frac{1}{2}a^2)f]' = \frac{1}{1 + z}\,,
\ee
whence
\be
(z^2 + \frac{1}{2}a^2)f = -\lambda\, z + c + \ln (1 + z)\,,\lb{Solf}
\ee
where $c = c(a^2)$ is an integration constant. Now we should show that this constant can be chosen in such a way that the right-hand side will be
${\cal O}(z^2 +  \frac{1}{2}a^2)$. We rewrite
\be
\ln (1 + z) = \frac{1}{2}\ln (1 - z^2) + \frac{1}{2}\left[\frac{\ln ( 1 + z)}{z} -  \frac{\ln ( 1 - z)}{z}\right]z\,\lb{decomp}
\ee
Both the first term and the expression inside the square brackets in \p{decomp} are functions of $z^2$ and we denote
\be
\frac{\ln ( 1 + z)}{z} -  \frac{\ln ( 1 - z)}{z} := g(z^2) = 2 + \frac{2}{3}z^2 + \frac{2}{5}z^4 + ...\,.
\ee
Now we denote $z^2 + \frac{1}{2}a^2 := y$ and rewrite
\be
\ln (1 - z^2) = \ln \left( 1 + \frac{1}{2}a^2\right) + \ln \left(1 - \frac{y}{1 + \frac{1}{2}a^2}\right)\,, \quad g(z^2) = g (y - \frac{1}{2}a^2)\,.
\ee
We wish the r.h.s. in \p{Solf} to start with $y$. This implies, first,
\be
c = - \frac{1}{2}\ln \left( 1 + \frac{1}{2}a^2\right)\,.
\ee
Second, we need to cancel the contribution from $g(-\frac{1}{2}a^2)$ which can be written as
\be
 \frac{1}{2i\sqrt{\frac{a^2}{2}}}\left[\ln \left( 1 + i\sqrt{\frac{a^2}{2}}\right) -  \ln \left( 1 - i\sqrt{\frac{a^2}{2}}\right)\right] z \lb{gcontrib}
\ee
Using the well known formula
$$
\arctan x = \frac{1}{2i}\ln \frac{1 +ix}{1 -ix}\,,
$$
we observe that \p{gcontrib} exactly cancels the $\lambda$ term in \p{Solf}. So we can divide both sides of \p{Solf} by $y$ and write
the nonsingular solution for the function $f$ as
\be
f(a^2,z) = \frac{\ln \left(1 - \frac{y}{1 + \frac{1}{2}a^2}\right)}{y} + \frac{1}{2y}\,\left[g(z^2) - g(-\frac{1}{2}a^2)\right].
\ee

\renewcommand\theequation{C.\arabic{equation}} \setcounter{equation}0
\section*{Appendix C\quad  }

In this Appendix we present some results of calculation of harmonic integrals depending on the fields $A^{ik}$ and $\bar A^{ik}$.
To compute the harmonic integrals
in \p{b-terms} and those appearing in the other pieces of the total bosonic action we apply the well-known Feynman formula
\be
\frac{1}{{\cal A}{}^p {\cal B}{}^q} = \int^1_0 dx \,\frac{x^{p-1}(1-x)^{q-1}}{[x{\cal A} + (1-x){\cal B}]^{p + q}}\,
\frac{\Gamma(p+q)}{\Gamma(p)\Gamma(q)}\,. \lb{FeDe}
\ee

We find:
\bea
\int du \frac{1}{(1 + A^{+-})(1 + \bar A^{+-})} &=& -\frac{1}{2\sqrt{\Delta}}\, \ln \frac{B_-}{B_+}\,,
 \lb{1int}\\
\int du \frac{1}{(1 + A^{+-})^2(1 + \bar A^{+-})} &=& \frac{1}{2\Delta} \Bigg[\frac{1}{2\sqrt{\Delta}}\,\big( \bar A^2 - A\cdot \bar A\big)\,
\ln \frac{B_-}{B_+} - 2\,\frac{A^2 - A\cdot \bar A}{2 + A^2}\Bigg], \lb{2int} \\
\int du \frac{1}{(1 + \bar A^{+-})^2(1 + A^{+-})} &=& \frac{1}{2\Delta} \Bigg[\frac{1}{2\sqrt{\Delta}}\,\big( A^2 - A\cdot \bar A\big)
\ln \frac{B_-}{B_+} - 2\,\frac{\bar A^2 - A\cdot \bar A}{2 + \bar A^2}\Bigg], \lb{3int}
\eea
\bea
\int du \frac{u^+_{(i}u^-_{k)}}{(1 + A^{+-})^2(1 + \bar A^{+-})} &=& \frac{1}{2\Delta}\,A_{ik} \Bigg( \frac{2 + A\cdot \bar A}{ 2 + A^2} +
\frac{ 2 + \bar A^2}{4\sqrt{\Delta}}\,\ln \frac{B_-}{B_+}\Bigg) \nn
 &&- \,\frac{1}{2\Delta}\,\bar A_{ik} \Bigg(1 + \frac{ 2 + A\cdot \bar A}{4\sqrt{\Delta}}\,\ln \frac{B_-}{B_+}\Bigg), \lb{4int} \\
\int du \frac{u^+_{(i}u^-_{k)}}{(1 + \bar A^{+-})^2(1 + A^{+-})} &=& \frac{1}{2\Delta}\,\bar A_{ik}
\Bigg( \frac{2 + A\cdot \bar A}{ 2 + \bar A^2} +
\frac{ 2 + A^2}{4\sqrt{\Delta}}\,\ln \frac{B_-}{B_+}\Bigg) \nn
 &&- \,\frac{1}{2\Delta}\,A_{ik} \Bigg(1 + \frac{ 2 + A\cdot \bar A}{4\sqrt{\Delta}}\,\ln \frac{B_-}{B_+}\Bigg). \lb{5int}
\eea
 In these formulas,
 \bea
B_\pm := 2 + A\cdot \bar A \pm 2\alpha_0\,\bar f^if_i\,, \quad \Delta := \alpha_0^2 (\bar f^i f_i)^2\,.
 \eea
\vspace{0.3cm}

\noindent{{\it The $f^i = \bar f^i = 0$ and $a^{ik} = 0$ limits.}\\

In the limit $f^i = \bar f^i = 0, A^{ik} = \bar A^{ik} = a^{ik}$ the expressions above go over to the  corresponding expressions from Appendix A.
Less trivial is the limiting case $a^{ik} = 0\,, A^{ik} = -\bar A^{ik} = ib^{ik} = -\alpha_0 \bar f^{(i}f^{k)}\,$. In this limit
\bea
&& {\rm (B.1)} \; \Rightarrow
\; -\frac{1}{\sqrt{\Delta}} \;\ln \frac{1 -\frac{\alpha_0}{2}(\bar f f)}{1 +\frac{\alpha_0}{2}(\bar f f)}\,, \nn
&&  {\rm (B.2)}  = {\rm (B.3)}\; \Rightarrow \;-\frac{1}{2\sqrt{\Delta}} \;
\ln \frac{1 -\frac{\alpha_0}{2}(\bar f f)}{1 +\frac{\alpha_0}{2}(\bar f f)} + \frac{1}{2}\,\frac{1}{1 - \frac{\alpha_0^2}{4}(\bar f f)^2}\,, \nn
&& {\rm (B.4)} = - {\rm (B.5)} \; \Rightarrow \; - ib_{ik}\,\frac{1}{\Delta} \Bigg[ \frac{1}{1 - \frac{\alpha_0^2}{4}(\bar f f)^2}
+ \frac{1}{\sqrt{\Delta}} \;\ln \frac{1 -\frac{\alpha_0}{2}(\bar f f)}{1 +\frac{\alpha_0}{2}(\bar f f)}\Bigg]. \nonumber
\eea

\end{document}